\documentclass[a4paper,fleqn,usenatbib,useAMS]{mnras}

\usepackage{graphicx}	

\usepackage{float}
\usepackage{subfigure}

\usepackage{amsmath}	
\usepackage{amssymb}	
\usepackage{multicol}        
\usepackage{bm}
\usepackage{pdflscape}	

\graphicspath{{./Figs/}}

\usepackage{enumitem}
\setlist{
  itemsep=10pt,
  parsep=10pt,
  topsep=10pt,
  listparindent=0pt,
  itemindent=0pt,
  leftmargin=\parindent
}

\usepackage{color}

\newcommand{\HI}{\hbox{\rmfamily H\,{\textsc i}}}
\newcommand{\HIsub}{\hbox{{\scriptsize H}\,{\tiny I}}}
\newcommand{\MHI}{\hbox{$M_{\HIsub}$}}

\newcommand{\DHI}{\hbox{$D_{\HIsub}$}}
\newcommand{\RHI}{\hbox{$R_{\HIsub}$}}
\newcommand{\Msun}{\hbox{${\rm M}_{\odot}$}}

\usepackage[T1]{fontenc}
\usepackage{ae,aecompl}

\usepackage{txfonts}

\title[Cubelet Stacking] {Interferometric Cubelet Stacking to Recover  H\,\textsc{i} Emission from Distant Galaxies}

\author[Qingxiang Chen et al.]{
Qingxiang Chen,$^{1,3}$\thanks{Contact e-mail: chenqingxiangcn@gmail.com}
Martin Meyer,$^{1,2}$
Attila Popping,$^{1,3}$
Lister Staveley-Smith$^{1,2}$
\\
$^{1}$International Centre for Radio Astronomy Research (ICRAR), University of Western Australia, 35 Stirling Hwy, Crawley, WA 6009, Australia\\
$^{2}$ARC Centre of Excellence for All Sky Astrophysics in 3 Dimensions (ASTRO 3D) \\
$^{3}$ARC Centre of Excellence for All-sky Astrophysics (CAASTRO)
}

\date{Last updated 2021 Jan 15}

\pubyear{2021}

\begin{document}
\label{firstpage}
\pagerange{\pageref{firstpage}--\pageref{lastpage}}
\maketitle

\begin{abstract}
In this paper we introduce a method for stacking data cubelets extracted from interferometric surveys of galaxies in the redshifted 21-cm H\,\textsc{i} line. Unlike the traditional spectral stacking technique, which stacks one-dimensional spectra extracted from data cubes, we examine a method based on image domain stacks which makes deconvolution possible. To test the validity of this assumption, we mock a sample of 3622 equatorial galaxies extracted from the GAMA survey, recently imaged as part of a DINGO-VLA project. 
We first examine the accuracy of the method using a noise-free simulation and note that the stacked image and flux estimation are dramatically improved compared to traditional stacking. The extracted H\,\textsc{i} mass from the deconvolved image agrees with the average input mass to within 3\%. However, with traditional spectral stacking, the derived H\,\textsc{i} is incorrect by greater than a factor of 2.
For a more realistic case of a stack with finite S/N, we also produced 20 different noise realisations to closely mimic the properties of the DINGO-VLA interferometric survey. We recovered the predicted average H\,\textsc{i} mass to within $\sim$4\%. 
Compared with traditional spectral stacking, this technique extends the range of science applications where stacking can be used, and is especially useful for characterizing the emission from extended sources with interferometers.
\end{abstract}




\section{Introduction}

Studying the properties of neutral hydrogen (H\,\textsc{i}) in galaxies is crucial to understanding how they form and evolve.  Hydrogen is the most abundant element in the Universe, and tracing the manner in which it feeds into galactic potential wells through accretion and interaction, and tracing the manner in which it gravitationally collapses into cold molecular clouds and stellar systems, is important in understanding the structure and dynamics of galaxies. However, H\,\textsc{i} can be difficult to detect observationally, and only a number of emission- and absorption-line methods are available.

At the highest redshifts, Lyman-$\alpha$ absorption has been the main tool for studying H\,\textsc{i}. Damped Lyman-$\alpha$ (DLA) systems in particular appear to trace the bulk of the cosmic H\,\textsc{i} density at these redshifts 
\citep{Lanzetta:1991a,Prochaska:2005a,Noterdaeme:2009a,Noterdaeme:2012a,Songaila:2010a,Zafar:2013a,Crighton:2015a,Neeleman:2016a,Bird:2016a,Rao:2006a,Rao:2017a}.
At redshifts less than about 1.6, where Lyman-$\alpha$ remains in the ultra-violet, expensive space telescope observations are required.
In the nearby Universe, the 21-cm hyperfine emission line has instead generally been preferred and, under the optical thin assumption, provides a direct method of tracing the distribution and morphology of H\,\textsc{i} gas, measuring its mass, and providing information on redshift and internal dynamics. 21-cm absorption features have also been used to detect neutral gas in distant galaxies, and H\,\textsc{i} narrow self-absorption (HINSA) has been used to determine optical depth in very nearby galaxies \citep{Li:2003a}.

During the last two decades, large blind surveys such as HIPASS \citep{Barnes:2001a,Meyer:2004a,Wong:2006a} and ALFALFA \citep{Giovanelli:2005a,Haynes:2018a} have advanced our understanding of H\,\textsc{i} gas in the Local Universe by observing large sky areas. Beyond the local Universe, a number of large surveys have also been conducted \citep{Catinella:2008a,Zwaan:2001a,Verheijen:2007a}, though mostly pre-selecting targets likely to have sufficient signal-to-noise (S/N) ratio.  A small number of blind surveys beyond the local Universe have also been carried out, reaching the limits of current telescope sensitivities and achievable integration times (e.g.\, AUDS: \citealt{Freudling:2011a,Hoppmann:2015a}, and CHILES: \citealt{Fernandez:2013a,Fernandez:2016a}).

To push beyond the limitations of direct detection studies, H\,\textsc{i} stacking is one of the principal methods that has been utilised to increase the redshift range over which emission-line measurements can be carried out. By using this method to determine the average {\HI} mass of the sample being stacked, it has been possible to discern the average H\,\textsc{i} properties of large samples and their dependence on other galactic constituents and environmental factors. For example, \citet{Fabello:2011a} studied the gas properties of low-mass H\,\textsc{i}-deficient galaxies in the nearby Universe. \citet{Fabello:2012a}, \citet{Brown:2015a}, and \citet{Brown:2017a} studied the relationship with environment. \citet{Fabello:2011b} and \citet{Gereb:2013a} studied the relationship between H\,\textsc{i} gas and AGN. \citet{Verheijen:2007a} and \citet{Lah:2009a} stacked spectra to study gas content in cluster environments at redshift up to 0.37.  
Additionally, comparison with optical luminosity functions can provide measurements of the cosmic H\,\textsc{i} density $\Omega_{\rm H\,\textsc{i}}$ \citep{Lah:2007a,Delhaize:2013a,Rhee:2013a,Rhee:2016a,Rhee:2018a,Kanekar:2016a}. Continuum based {\HI} stacking is also used for star formation rate measurements \citep[e.g.][]{Bera:2018a,Bera:2019a}. These studies have shown the utility of {\HI} stacking to probe the statistical properties of galaxies up to redshift unity and beyond.


Whilst the next generation of radio facilities, such as ASKAP \citep{Johnston:2008a,Meyer:2009a}, MeerKAT \citep{Holwerda:2012a}, FAST \citep{Nan:2011a,Duffy:2008a}, and WSRT/APERTIF \citep{Oosterloo:2009a} will enable large area deep blind surveys and result in many direct detections, H\,\textsc{i} stacking will allow the redshift limits of H\,\textsc{i} analysis to be pushed further still in cases where suitably large optical redshift surveys exist.  This will be particularly important for studying the gas content of the Universe at higher redshifts, and studying the processes behind the rapid rise and decline of the star-formation rate density \citep{Hopkins:2006a,Madau:2014a}.


A significant difference between upcoming SKA pathfinder studies, and many of those that have been carried out to date, is that future work will primarily be carried out using interferometers, compared to the single dish studies that have dominated existing work. This offers both significant advantages as well as new challenges. A major advantage of interferometers is that confusion can be overcome, which is otherwise a significant limitation in single-dish studies. Confusion effects in forthcoming H\,\textsc{i} surveys have been investigated by \citet{Duffy:2012b, Duffy:2012a} and \citet{Jones:2015a, Jones:2016a}. \citet{Jones:2016a}  estimated the effect of confusion in H\,\textsc{i} stacking. With a synthesis beam size of $\sim 10''$, confusion only has a minor effect on the stacked H\,\textsc{i} mass for the redshift range of surveys considered. Interferometers, such as ASKAP and MeerKAT, will be able to significantly reduce the impact of confusion with their higher angular resolutions while the SKA, with its additional sensitivity, will be capable of extending studies over a large fraction of cosmic time, and reduce confusion effects even further.

However, there remain two significant challenges for interferometric stacking.  First, the H\,\textsc{i} signals from individual galaxy observations are generally too weak detect directly.  As such, deconvolution is not possible as the S/N is too low. So traditional spectral stacking experiments are effectively conducted on dirty images. For spatially unresolved sources this does not pose a problem. In a point source sample without confusion, the initial flux unit Jy/beam can be simply replaced by Jy, as all flux is enclosed in a single synthesised beam. Flux can therefore be simply extracted using the central pixel from the dirty image \citep{Meyer:2017a} without considering the pixels affected by sidelobes. Indeed, some experiments deliberately degrade telescope resolution so that this is not a problem.  However, for spatially resolved observations, the final stacked H\,\textsc{i} spectra will be subject to sidelobe contamination and therefore incorrect.
This will lead to imprecise estimates of measured quantities, including mass, extent, dynamics, and structure.
Second, the problem of poor uv-coverage for many interferometers means that there will invariably be major differences between `dirty' images and true images, which compounds the issue of studying low S/N sources, compromising the ability to estimate the true flux of extended, or even marginally extended sources. This can be mitigated by using multi-configuration observations, utilising earth rotation synthesis, and applying optimal weighting in order to achieve a better Point Spread Function (PSF). However, this requires long integration times, and reduces the area of sky that can be covered in an experiment. Moreover, most interferometers are simply too sparse to achieve anything like an ideal Gaussian PSF without using deconvolution techniques.

In this work, we introduce a new strategy for H\,\textsc{i} stacking: the cubelet stacking method. Instead of extracting each spectrum from dirty images and then stacking the 1D spectra, we stack the dirty 3D cubelets of galaxies (small data cube patches cut out from the original dataset), and then extract the stacked spectra. This has all the advantages spectral stacking has, while at the same time avoiding the two shortcomings mentioned above. This method makes it possible to deconvolve the combined cubelet, as the S/N of the combined cubelet is much higher than the individual cubelets. Additionally, the stacked PSF shape is often substantially better than the PSFs for the individual pointings, due to the varying uv-coverage as a function of time, frequency and pointing direction.


In this paper, we demonstrate the method, and test its suitability using a simulated survey. Section 2 introduces the DINGO VLA survey on which the simulation is based. Section 3 describes our cubelet stacking method. In Section 4, we use scaling relations to create the simulated DINGO VLA data, and proceed to test the stacking method. The results and discussion are finally presented in Sections 5 and 6.  A 737 cosmology is used throughout, that is, $\rm H_0=70$ km\ s$^{-1}$ Mpc$^{-1}$, $\Omega_{\rm M}=0.3$ and $\Omega_\Lambda = 0.7$.

\section{Data}

The simulated data used in this paper is based on the H\,\textsc{i} pathfinder observations taken for the Deep Investigations of Neutral Gas Origins survey (DINGO) on the Jansky Very Large Array (VLA), in combination with optical redshifts data from the Galaxy and Mass Assembly survey (GAMA)\footnote{www.gama-survey.org}. This dataset is briefly introduced here, but described in more detail in Chen et al. (prep).  The comparatively shallow nature of the VLA observations, along with their sparse equatorial uv coverage, make this an ideal dataset to refine and test the cubelet stacking methodology.

The VLA data were observed in semesters 2014B and 2016A.  In total, 270 pointings were observed, with a total sky area of $\sim$40~deg$^2$. The full width half maximum (FWHM) primary beamwidth of the VLA antennas in the redshift range examined ($z<0.1$) is $\sim$30$\arcmin$.  Three pointings were grouped together in each observing block over a 2 hour period, which included primary and secondary calibrator observations. Each pointing has $\sim$28~min  of observing time per 2 hour block. The uv-coverage in each pointing is relatively sparse, compared to a full earth rotation synthesis observations. The PSF therefore has significant non-Gaussian features (see Fig~\ref{fig:typical_psf}).  The data were processed with {\sc casa} standard data reduction tasks, including flagging, calibration and imaging (see Chen et al. (prep)). After the data reduction steps, three fields were completely flagged because of poor data quality.  This results in a final dataset of 267 spectral line cubes of $1024\times 1024\times\ 2048$ pixels, each cell being a size of $\rm 2\arcsec\times 2\arcsec\times 62.5 kHz$. The simulated PSFs were assumed to be identical to the PSFs calculated from the above uv-coverage, after flagging. No signal data is used in this paper, but noise from signal-free portions of the data is used to conduct the simulation.

Optical redshift data for the simulation were based on the internal version of GAMA DR3 catalogue \citep{Baldry:2018a}, choosing only those redshifts that overlap with the frequency range of the VLA data cubes.  Only galaxies with redshift robustness $\rm NQ>2$ are selected. After cross-matching, together with additional considerations (the galaxy not being too close to the frequency or spatial edges of data cubes, and the galaxy being within the primary beam), this results in an input stacking catalogue of 3622  galaxies. Within the 267 VLA pointings, these 3622 galaxies result in 5442 independent spectra and image cubelets, as some of the galaxies are located at edge of several pointings and hence observed more than once.

\section{Stacking Methodology}

In this section, we summarize the methodology of the traditional H\,\textsc{i} spectral stacking method and introduce the technique of cubelet stacking. For normal spectral stacking, individual spectra are first extracted from data cubes based on their optical 3D coordinates (i.e.\ position and redshift). Then a weighted averaging process is applied to these spectra. The weights can take into account the noise in the data, the distance from the telescope pointing centre and the distance of the galaxies. Using the final stacked spectrum, the average properties of the galaxy sample, such as flux, H\,\textsc{i} mass and linewidth can be studied. Cubelet stacking works in a similar way, with stacking happening in the image rather than the spectral domain - i.e.\ spectra from a range of positions around the central position of the galaxies are stacked. However, the PSF cubelets are also extracted and stacked using an identical weighting procedure. The main benefit of this method is that, when the stacked image cube has significant S/N ratio, it becomes possible to deconvolve the stacked cube which, as shown in this paper, results in more accurate image stacks, and flux and mass measurements for resolved galaxies.

\subsection{Spectral Stacking}

From an input optical catalogue with positions and redshifts, we define a sample of size $N$, with subscript $l=1,2,...,N$ indicating the individual spectra extracted. For a given $l$-th spectrum, we use the subscript $k$ to indicate different frequency channels. The series of flux densities is denoted $s_{k,l}$, and the series of corresponding frequencies is denoted $f_{k,l}$. The aperture size used for extracting the spectra is $A$. This can be an angular area, or physical area. Based on whether or not sources are resolved, the aperture can also be chosen as single pixel or an extended area. 

We first form the spectrum $s_{A,k,l}$ within $A$. The $s_{A,k,l}$ are in units of Jy or Jy/beam, depending on whether the data are spatially integrated or not. However, as noted later, the beam (PSF) size for a dirty image from an interferometer is not well defined.

We apply the following three steps for each spectrum before stacking:
\begin{enumerate}
\item{Blueshift the observed frequency back to the rest frame:
    $f'_{k,l} = f_{k,l} (1+z_l)$.
In order to conserve flux, we also apply the cosmological stretch factor:
$s'_{A,k,l} = s_{A,k,l} / (1+z_l)$.
}

\item {
The frequency axes from the different spectra are now misaligned, so we linearly interpolate each spectrum to a pre-defined frequency array ($f'_{k,l} \xrightarrow{interp} F_{k,l}$, $s'_{A,k,l} \xrightarrow{interp} s''_{A,k,l}$) prior to stacking. The channel spacing of the pre-defined array is 62.5 kHz.


}

\item {The flux densities for spectra not extracted from the centre of the primary beam need to be corrected for primary beam attenuation: $S_{A,k,l} =  s''_{A,k,l}/p_l$, where $p$ is the VLA primary beam response for a given galaxy offset.

}

\item {The rms noise for each channel in each spectrum
is calculated from regions in the corresponding image plane without any emission. 
The rms noise array is represented as $\sigma_{k,l}$. }

\end{enumerate}

At this point we stack the spectra, where the stacked spectrum $S'_{A,k}$ is:
\begin{equation}
S'_{A,k} = \frac{ \sum_l S_{A,k,l} w_{k,l} } {\sum_l w_{k,l}},
	\label{eq:y_stk_spec}
\end{equation}
where $w$ denotes the weight assigned to each spectrum and spectral channel. In this paper, we use weight factors which relate only to  primary-beam corrected noise levels and luminosity distances:
\begin{equation}
w_{k,l} = \sigma_{k,l}^\xi d_l^\gamma. 
	\label{eq:wt}
\end{equation}
We apply normal noise variance weighting ($\xi=-2$), but should also consider differing values for $\gamma$, as this choice balances S/N of the stack, the weighted mean redshift of the sample, and the cosmic volume probed. $\gamma = 0$ corresponds to uniform weighting \citep[e.g.][]{Delhaize:2013a,Rhee:2013a} whereas $\gamma = -1$ suppresses noise from high-redshift samples, while not over-biased by low-redshift galaxies \citep[e.g.][]{Hu:2019a}. For simplicity, throughout this paper we take $\gamma=0$.

Although we express the spectral stacking process in flux, other kinds of spectra (i.e. in mass, M/L) can be stacked in the same way. In this work we test the method using mass spectra, so a unit conversion is conducted before the stacking. The conversion between H\,\textsc{i} mass and flux is given by:
\begin{equation}
\left(\dfrac{ M_{\rm H\,\textsc{i}} } { {\rm M}_\odot\, {\rm beam}^{-1} }\right) = 49.7\left(\dfrac{D_L}{\rm Mpc}\right)^2\left(\dfrac{ F_{\rm H\,\textsc{i}}}{\rm Jy\, Hz\, beam^{-1}}\right),
	\label{eq:flux2mass}
\end{equation}
where $D_L$ is luminosity distance, and $F_{\rm H\,\textsc{i}}$ is spectrally integrated flux.

\subsection{Cubelet Stacking}

For every galaxy in this sample, we extract a small image cubelet from the full dirty cube, centered on its optical position and  redshift converted into a redshifted H\,\textsc{i} frequency. We also extract a same-sized PSF cubelet from the full PSF cube, centered on peak of the PSF. The size of extracted cubelets should be big enough to consists most of galaxy signals, and big enough to include main features of PSF, as a deconvolution process will be conducted later.

Subscripts for RA, Dec and channels are denoted by $i,j,k$, respectively; pixel values in the image cubelet are given by $x_{i,j,k}$, pixel values in PSF cubelet by $y_{i,j,k}$, and the frequency values in each cubelet are given by $f_k$. 
We apply the following four steps to the image and PSF cubelet of each galaxy before stacking:

\begin{enumerate}
\item {Blueshift the observed frequency axis to the rest frame, and conserve flux as before:
$    f'_{k,l} = f_{k,l} (1+z_l)$; 
%
$x'_{i,j,k,l} = x_{i,j,k,l} / (1+z_l)$.
}

\item{Linearly interpolate the frequency axis of the image and PSF cubelet to a pre-defined frequency array:
$f'_{k,l} \xrightarrow{interp} F_{k,l}$;
%
$x'_{i,j,k,l} \xrightarrow{interp} x''_{i,j,k,l}$;
%
$y_{i,j,k,l} \xrightarrow{interp} Y_{i,j,k,l}$.
}

\item {Apply primary beam correction:
%
%
$X_{i,j,k,l} = x''_{i,j,k,l} / p_l$,
where $p$ is the VLA primary beam response for a given galaxy offset.}


\item{
We determine the cubelet channel noise $\sigma_{k,l}$ in a similar manner as for the spectral stacking. The noise is derived using all image pixels at a given frequency channel, and not just those in the extracted region.
}

\end{enumerate}

After applying the above to all spectra in the sample, we stack both the image cubelets and PSF cubelets. The stacking steps are similar to spectral stacking, except we now average pixel values in the image domain and the spectral domain.
For the image cubelets, we calculate:
\begin{equation}
X'_{i,j,k} = \frac{ \sum_l X_{i,j,k,l} w_{k,l} } {\sum_l w_{k,l}}.
	\label{eq:x_stk}
\end{equation}
For PSF cubelets, we calculate:
\begin{equation}
Y'_{i,j,k} = \frac{ \sum_l Y_{i,j,k,l} w_{k,l} } {\sum_l w_{k,l}},
	\label{eq:y_stk}
\end{equation}
where the weight factors are the same as for spectral stacking.

Note the spectra stacked in these steps can be in flux, mass, M/L or other kinds. In this work we test our method using mass spectra. The conversion between flux spectrum to mass spectrum is in Eq.~\ref{eq:flux2mass}.


Following stacking, we can then implement deconvolution, whereby the stacked image cube $X'$ is {\sc clean}ed under H\"{o}gbom algorithm with the stacked PSF cube $Y'$.
From the deconvolved stacked image cube, a spectrum can then extracted from within a given aperture, and a total flux estimated, taking into account the restoring beam area. Taking the spatial aperture size as $A$, the clean image as $C$, and the Gaussian-fit to the restoring beam from the stacked PSF as $G$, we can approximate the spatially integrated spectrum by:
\begin{equation}
S_{A,k} = \frac{ \sum_A C_{i,j,k} } {\sum_A G_{i,j,k}}.
	\label{eq:spectrum}
\end{equation}

Note that $A$ should not exceed the central quarter in the RA-DEC domain, otherwise the {\sc clean} algorithm cannot deconvolve the PSF from the full cubelet.

\section{Simulation}

In this section we simulate H\,\textsc{i} spectral stacking and cubelet stacking to test our method. We use optical scaling relations to simulate the H\,\textsc{i} content of the observed galaxy sample and try to recover the simulated signal, i.e.\ average H\,\textsc{i} mass. In this simulation, we concentrate on stacking spectra scaled to mass density units, rather than flux density units.

We first interpolate magnitudes from the GAMA catalogue to $B$-band using the SDSS transformation model:\footnote{Thanks to Robert Lupton's work in 2005: \\ http://classic.sdss.org/dr4/algorithms/sdssUBVRITransform.html\#Lupton2005}
\begin{equation}
B = g + 0.3130\ (g - r) + 0.2271.
	\label{eq:filter_convert}
\end{equation}
Here $B$, $g$ and $r$ are all absolute magnitudes in the three bands. The absolute magnitude of $g$ and $r$ are calculated based on Petrosian magnitudes, distance moduli, dust extinctions and $k$-corrections provided from \citet{Baldry:2018a} and \citet{Loveday:2012a}.

\begin{figure}
    \includegraphics[width=\columnwidth]{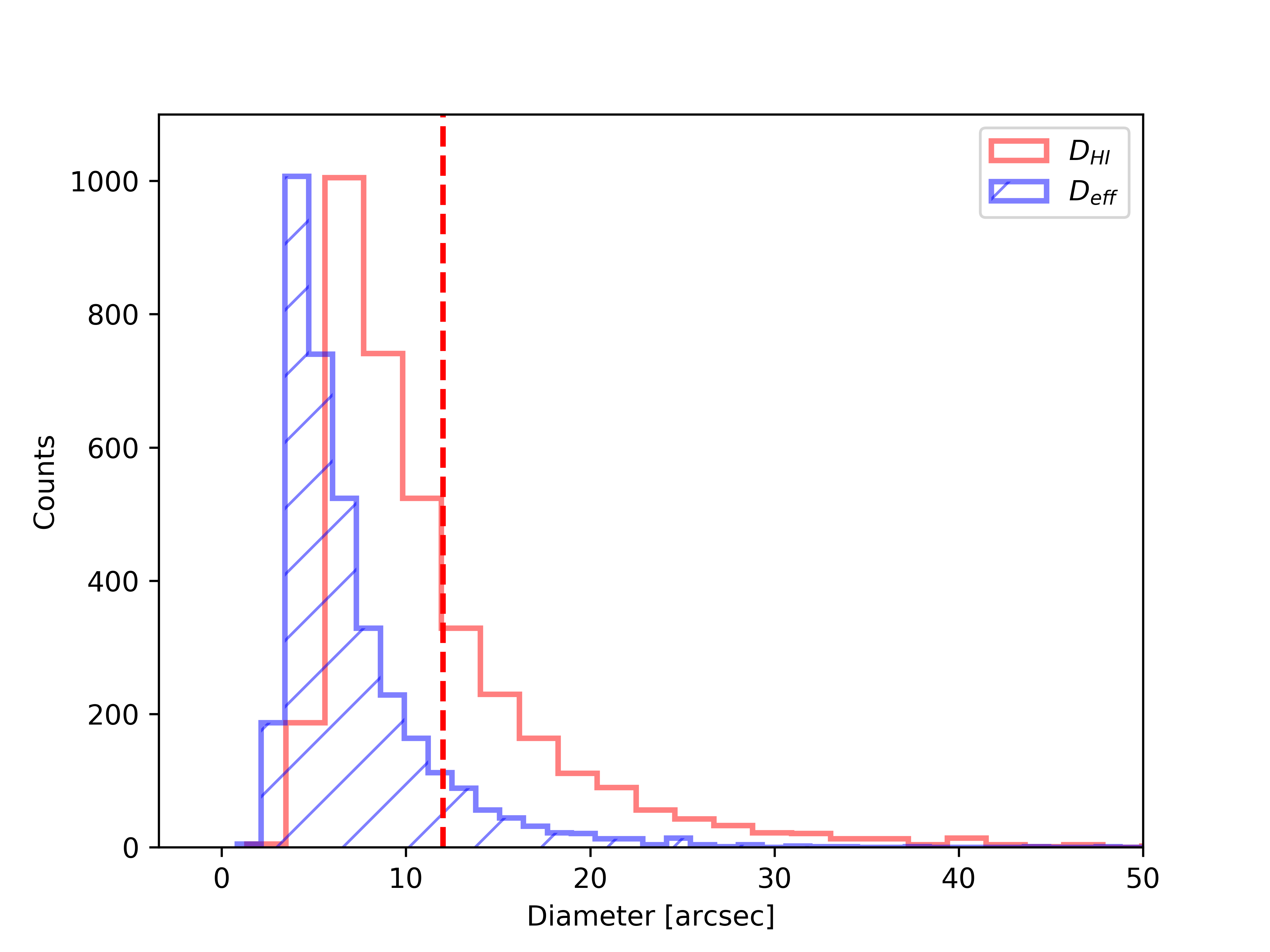}
    \caption{The distributions of H\,\textsc{i} galaxy diameters, $D_{\rm H\,\textsc{i}}$ and $D_{\rm eff}$ predicted from scaling relations, for the 3622 GAMA sample. The blue shaded region represents the histogram for H\,\textsc{i} effective diameter. The unfilled red histogram shows H\,\textsc{i} diameters defined by the position where the H\,\textsc{i} surface density equals 1 M$_{\odot}$ pc$^{-2}$. The vertical dashed line at 12$\arcsec$ represents the minor axis of a Gaussian fit to the stacked VLA PSF used in this simulation. 
    }
    \label{fig:scaling-relations}
\end{figure}

We then take the scaling relation from \citet{Denes:2014a} to predict approximate H\,\textsc{i} masses using:
%
\begin{equation}
\log M_{\rm H\,\textsc{i}} = 2.89 - 0.34\ B.
	\label{eq:sr_mass}
\end{equation}
Combining Eq.~\ref{eq:sr_mass} and the H\,\textsc{i} size scaling relations in B band from \citet{Broeils:1997a}, we derive:
%
\begin{equation}
\log D_{\rm H\,\textsc{i}} = 0.4921\log M_{\rm H\,\textsc{i}} -3.3766,
	\label{eq:sr_size}
\end{equation}
and
\begin{equation}
\log D_{\rm eff}= 0.4924\log M_{\rm H\,\textsc{i}} - 3.5918,
	\label{eq:sr_eff}
\end{equation}
where $D_{\rm H\,\textsc{i}}$ is the diameter in kpc where the surface mass density drops down to 1 M$_{\odot}$ pc$^{-2}$ and $D_{\rm eff}$ is the half-mass radius in units of kpc. The size distribution of the 3622 GAMA galaxy sample using scaling relations is shown in Fig~\ref{fig:scaling-relations} after converting these diameters to angular sizes. In this work we concentrate on the latter definition (Eq.~\ref{eq:sr_eff}) to represent galaxy H\,\textsc{i} sizes.

For every galaxy, we take $D_{\rm H\,\textsc{i}}$ calculated from scaling relations as the length of the major axis. We take the ellipticity $\eta$ from the $r$-band data to calculate the minor axis $b=D_{\rm H\,\textsc{i}} (1-\eta)$. We also take the $r$-band position angle into consideration when simulating the shape of the galaxies.
We then allocate the H\,\textsc{i} mass uniformly to all the voxels occupied by a galaxy, assuming every galaxy has a fixed line-width of 1 MHz. 
The size for every cubelet is 200 pixels $\times$ 200 pixels $\times$ 160 channels, with the pixel size being $\rm 2\arcsec\times2\arcsec\times62.5 kHz$. A line-width of 1 MHz corresponds to 16 channels in the simulation. Note that the allocated sample size is 5442 rather than 3622, in order to match the real observational sample.

\begin{figure}
    \includegraphics[width=\columnwidth]{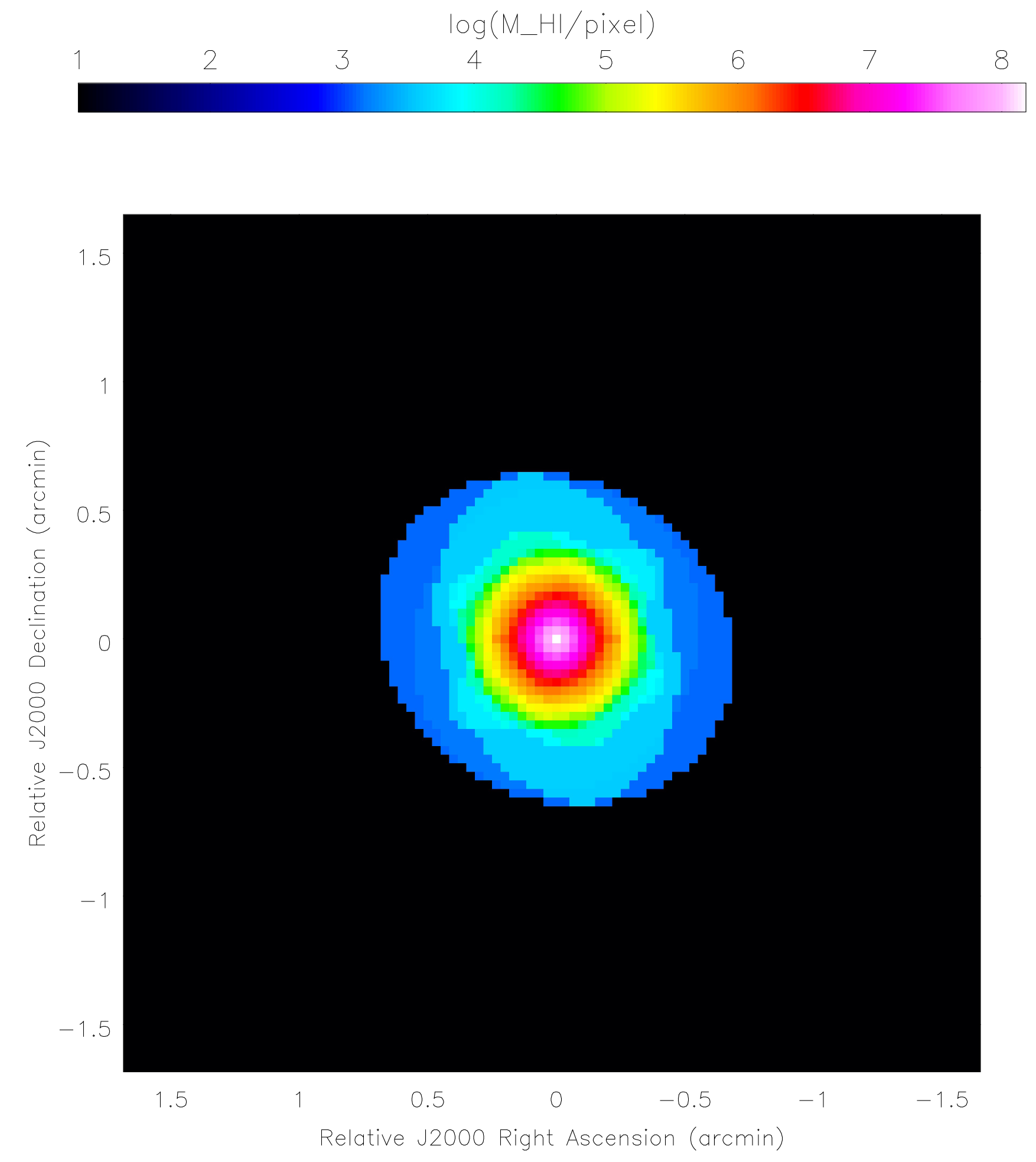}
    \caption{The stacked model image from the simulation. The model image for every galaxy is based on the H\,\textsc{i} mass predicted from the optical magnitudes and colour, and convolved with a top hat model based on the optical position angle and the H\,\textsc{i} size, also predicted from the optical magnitude. Noise-based weighting is used in the model stacking, although the image itself does not include noise. The stacked simulation has an H\,\textsc{i} mass of $1.74\times 10^9$ M$_\odot$. The input images have not been convolved with the PSF.
    }
    \label{fig:model_stk}
\end{figure}

\begin{figure}
    \includegraphics[width=\columnwidth]{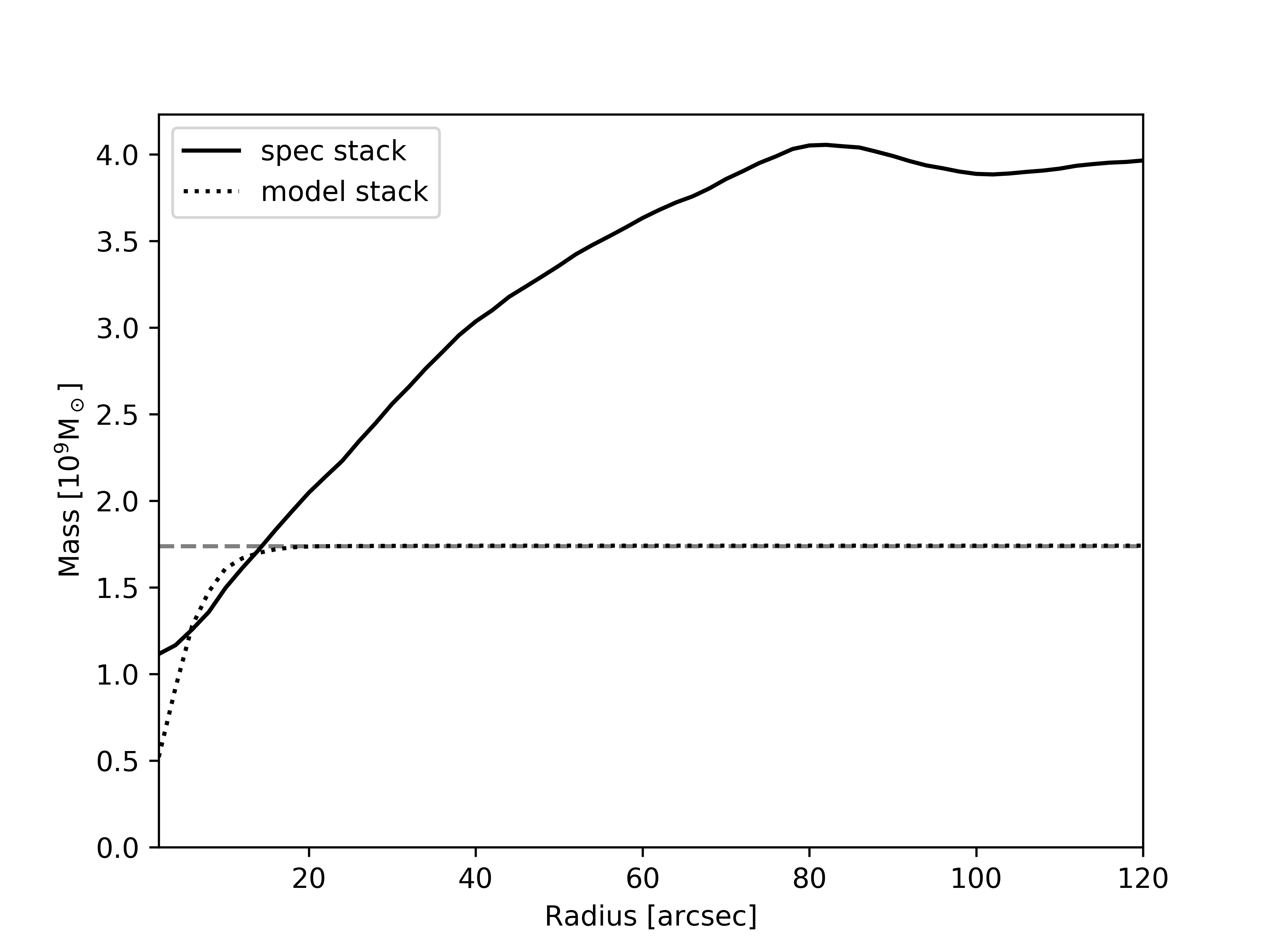}
    \caption{The cumulative H\,\textsc{i} mass of the stacked model and spectral stacking results, as function of aperture size. Of all the H\,\textsc{i} mass, 99.7\% is within 20$\arcsec$ radius. The solid, dotted and dashed lines indicate the spectral stacking measurement, model stacking measurement and expected values, respectively. Simple application of spectral stacking results in a discrepancy in the measured mass by a factor of 2.3 due to the nature of the psf.}
    \label{fig:model_profile}
\end{figure}

We then stack the 5442 models and the PSFs
using realistic weight factors $w_{l,k}=\sigma_{l,k} ^{-2}$, even in the noise-free simulation. The noise information comes from the observed data cubes. This makes it consistent with further simulations with noise added.  As motioned before, we do not consider distance weightings for simplicity. The stacked model image is shown as Fig~\ref{fig:model_stk}. The stacked H\,\textsc{i} mass is $\rm 1.74\times 10^9M_\odot$. 
The curve of growth for the stacked H\,\textsc{i} mass distribution (Fig~\ref{fig:model_profile}) shows that 99.7\% of the H\,\textsc{i} mass is located within a radius of 20$\arcsec$.

The next step is to stack the convolved model image cubes, with and without noise, to test whether the above model measurements can be reproduced.


\subsection{Spectral Stacking}
\label{sec:no-noise-spec}

In this test, we first convolve the extended models with their actual observed PSF, to mock an noise free observation. We then stack spectra extracted from these convolved images. Pixel units are converted from Jy~beam$^{-1}$ to M$_\odot$ beam$^{-1}$ through Eq.~\ref{eq:flux2mass}. A 2D Gaussian function is fit to each PSF to estimate the beam shape, and spectra are extracted by summing up the pixel values in circular area $A$ from the convolved images. The recovered mass spectrum for each galaxy is calculated using:
\begin{equation}
\frac{M_{A,k}}{\rm M_\odot} = \frac{M'_{A,k}  G_{A,k}^{-1}}{\rm M_\odot~beam^{-1}} .
	\label{eq:gauss_convert}
\end{equation}
In this equation, $M'_{A,k} $ is the spectrum extracted from the simulated image; $G_{A,k}$ is from summing up the values in area $A$, of the corresponding Gaussian PSF approximation normalised to unity at the central pixel. The extracted spectra are then stacked using Eq.~\ref{eq:x_stk}. The final simulated H\,\textsc{i} mass is then derived by integrating the emission from the central 16 channels.

\subsection{Cubelet Stacking}

\subsubsection{Noise-free cubelet stacking}
\label{sec:no-noise-cube}

For cubelet stacking, we stack in the image domain using Eq.~\ref{eq:x_stk}, and generate the stacked PSF image cubelet via Eq.~\ref{eq:y_stk}. After stacking the convolved model images, we deconvolve with the stacked PSF. We use a H\"{o}gbom {\sc clean} with a gain of 0.1. We apply 20, 50, 100, 200, and 500 iterations respectively to examine {\sc clean} convergence.

\subsubsection{Cubelet stacking with noise}
\label{sec:noise-cube}

We add noise to the simulated cubelets in order to better mimic the situation of real observations. For each of the 267 observed pointings, we randomly choose 100 3D positions where no source can be found in the GAMA catalogue within a distance of 200 pixels $\times$ 200 pixels $\times$ 96 channels ($400\arcsec\times 400\arcsec \times$ 6 MHz). We then extract a noise cubelet of size $400\arcsec\times400\arcsec\times$ 10 MHz. 
We thus obtain 26,700 noise cubelets from the 267 fields. 
We then randomly allocate these noise cubelets to the 5442 image cubelets, repeating the process 20 times,  giving 20 noise realisations for each model cubelet. The noise cubelets are then added to the convolved model images, and an identical process of blueshifting, interpolation, primary beam correction, conversion to mass units, and stacking is followed.


Finally, we deconvolve the stacked image cubelet with the stacked PSF cubelet. Signal-free channels from the stacked image cubelet are used to calculate the $rms$ level, which is then set as the {\sc clean}ing threshold unit. The integrated H\,\textsc{i} mass is calculated by integrating within the central 16 signal channels.

\section{Results}

\subsection{Spectral Stacking}

The relationship between the measured H\,\textsc{i} mass and aperture in the case of simple interferometric spectral stacking is shown as the solid line in Fig~\ref{fig:model_profile}. The measured mass reaches a constant value at the relatively large radius of $\sim$80$\arcsec$. This is much larger than the expected H\,\textsc{i} galaxy sizes -- almost all the  gas is expected to be located within an aperture of $20\arcsec$ (see Figs~\ref{fig:model_stk} and \ref{fig:model_profile}). The maximum value is $\rm 4.06\times 10^9 M_\odot$ at $\sim$82$\arcsec$ radius while the predicted value from the scaling relation models is $\rm 1.74 \times 10^9 M_\odot$. Thus, simple interferometric spectral stacking gives a stacked H\,\textsc{i} mass which is overestimated by a factor of 2.3. 

This discrepancy arises from the combination of non-Gaussian beams, poor uv-coverage and extended emission. Since the spectral stacks are not deconvolved, spatial integration will give an incorrect integral when the normal assumption of Gaussian beam is made.
The trend for H\,\textsc{i} mass to continue to increase at $R>20\arcsec$ is due to sidelobes in two ways. First, there is no {\sc clean} step, so sidelobes contaminate the measured curve. Second, as can be seen from Fig~\ref{fig:typical_psf}, the sidelobes extend to a large radii even at the 10\% level.

This implies, that for interferometric stacking of spectral data, the wrong value for H\,\textsc{i} mass will be obtained if the interferometer resolves the galaxies being observed, unless the interferometer beam is very clean (for example Gaussian). The predicted VLA beam in this simulation is highly non-Gaussian, partly reflecting the nature of the VLA baseline distribution, but not helped by the equatorial placement of the field being simulated.
Fig~\ref{fig:typical_psf} is a typical PSF image, from the central channel of PSF cube for field 231 (one pointing out of the 267 simulated). 
As extensively discussed in \citet{Jorsater:1995a}, the flux of an extended source can be easily overestimated by a large factor if it is measured directly from dirty images. This is because of poor \textit{uv}-coverage leading to highly non-Gaussian PSFs.

\subsection{Cubelet Stacking}

\begin{figure*} 
	\centering
	\vspace{0cm}
	\subfigtopskip=0pt 
	\subfigbottomskip=0pt 
	\subfigcapskip=0pt
	\subfigure[Stacked PSF]{
		\label{fig:stk_psf}
		\includegraphics[width=0.45\linewidth]{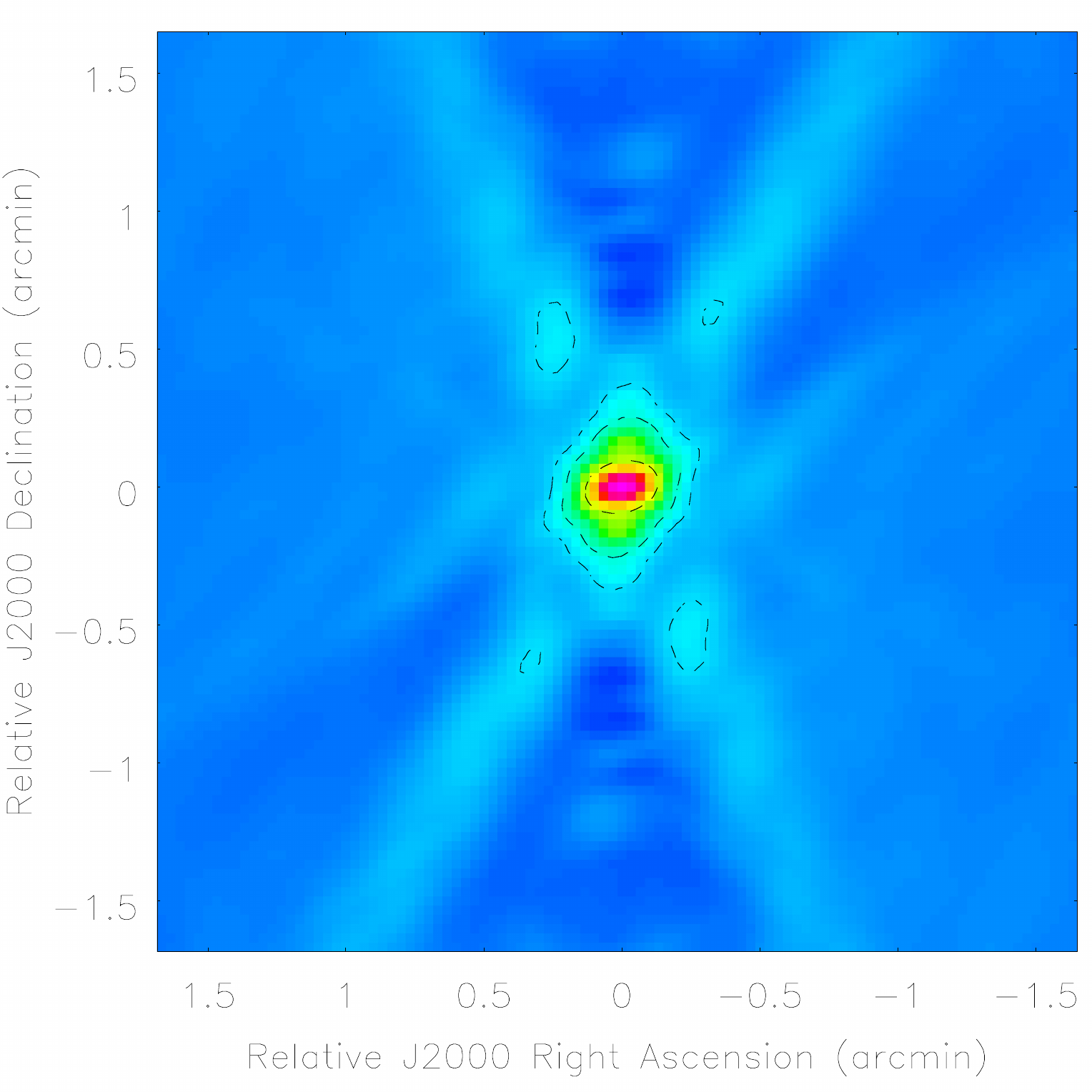}}
	\subfigure[Typical PSF without stacking]{
		\label{fig:typical_psf}
		\includegraphics[width=0.45\linewidth]{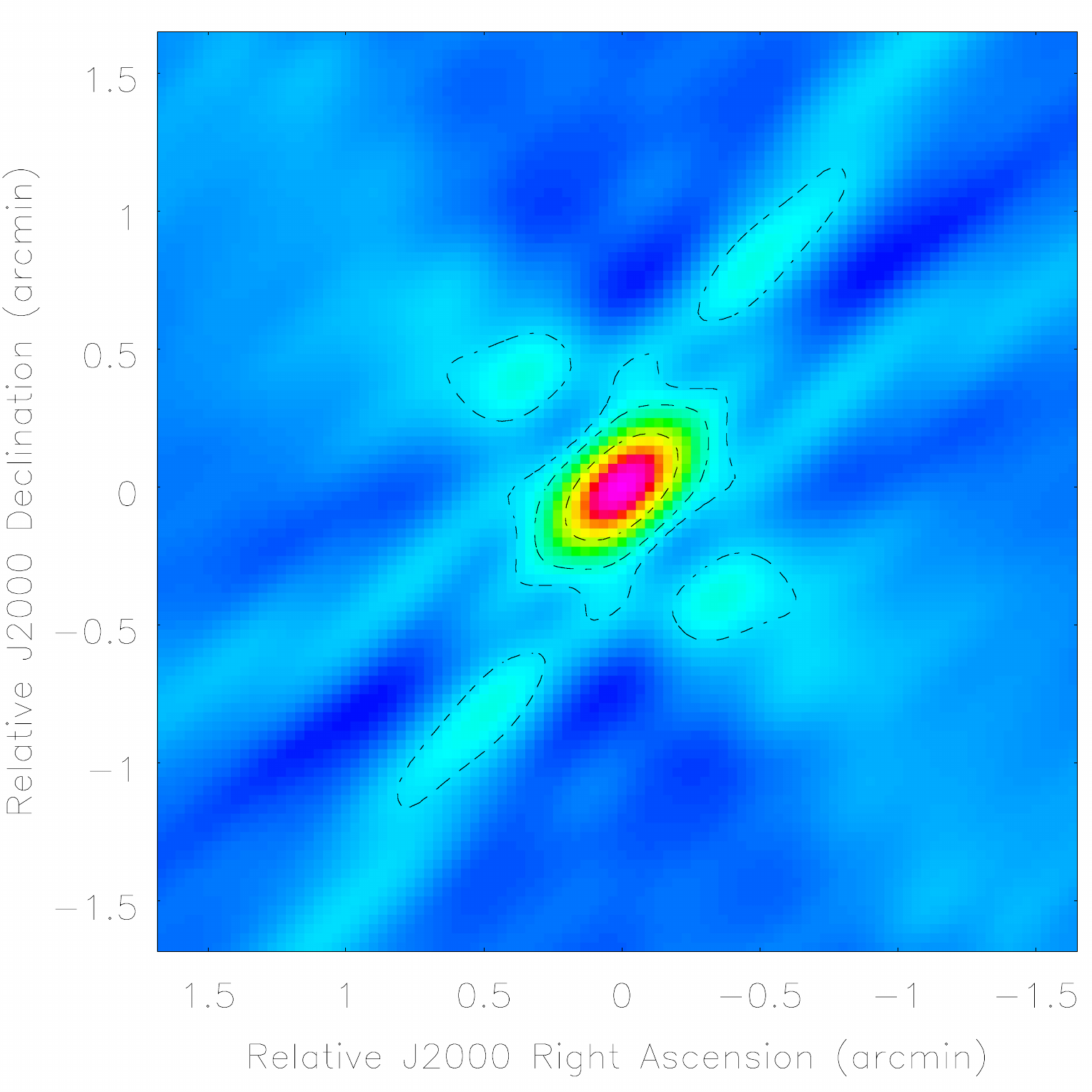}}
	\subfigure[Stacked dirty image]{
		\label{fig:stk_dirty}
		\includegraphics[width=0.45\linewidth]{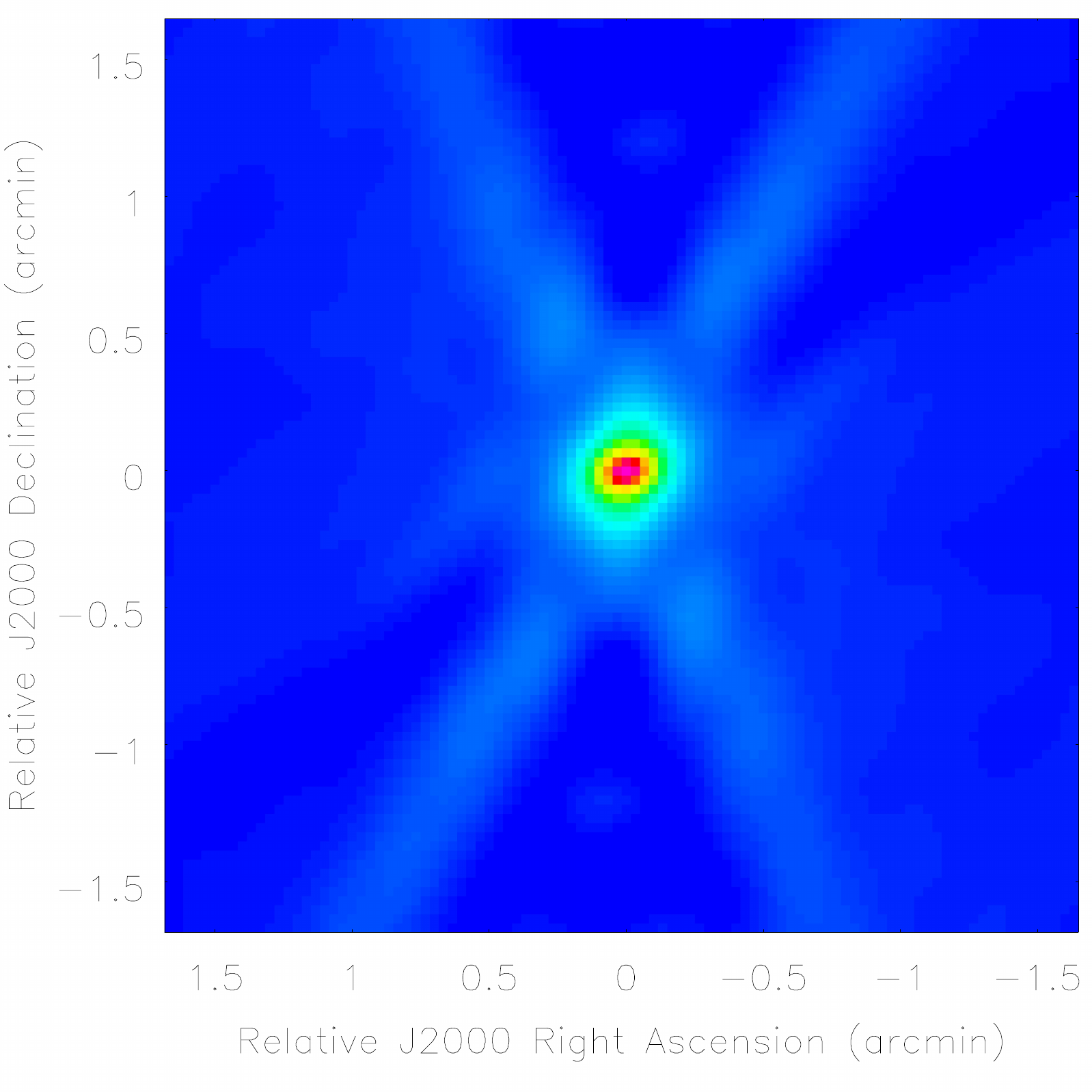}}
	\subfigure[Stacked image after deconvolution]{
		\label{fig:stk_decon}
		\includegraphics[width=0.45\linewidth]{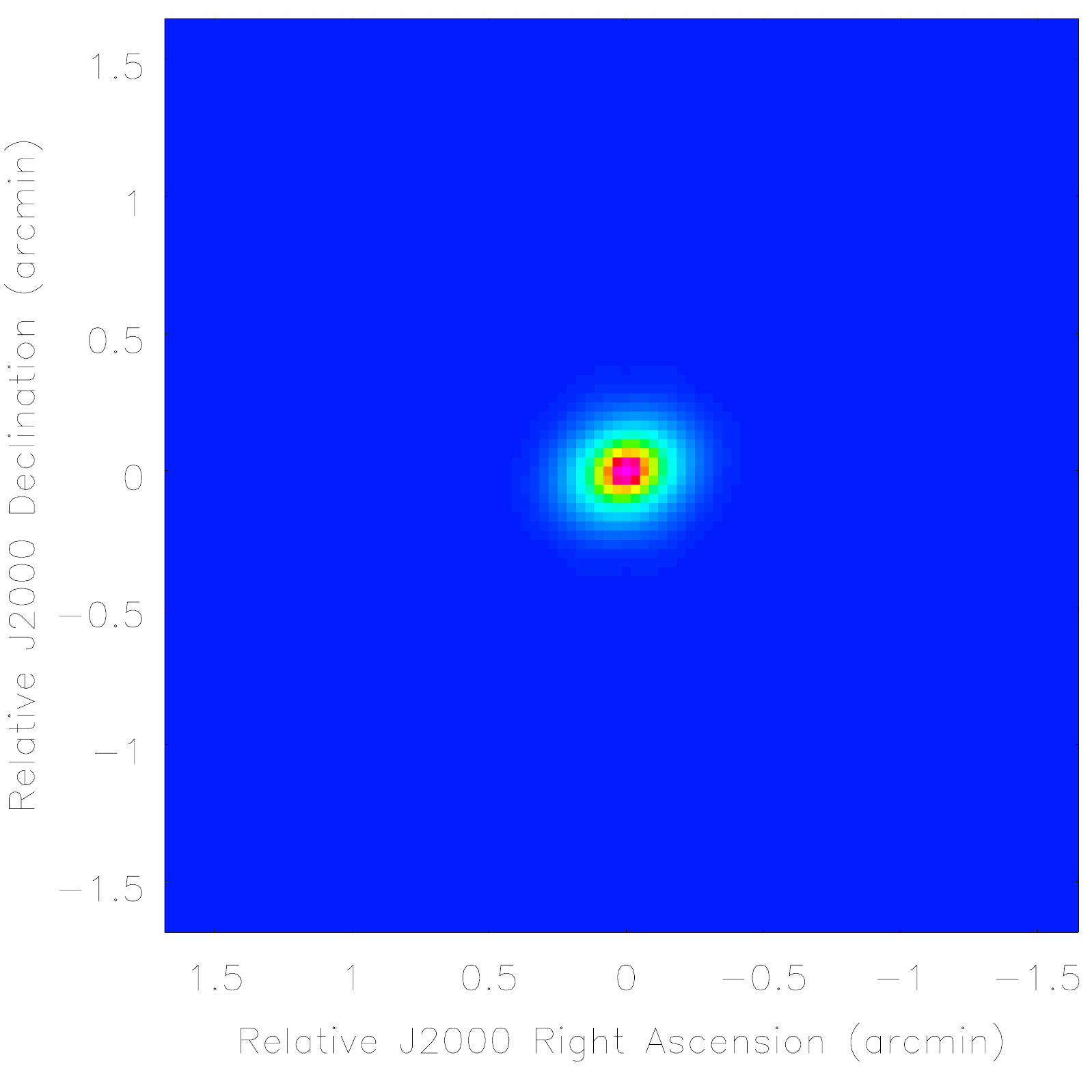}}
	\caption{\textit{Upper:} the PSF before and after stacking. (a): the stacked rest-frame PSF for 5442 individual PSFs. (b): a PSF image from a typical observation pointing (field 231) at the central channel. The colour scales of the two images are identical. The contour levels from the centre outwards are 0.5, 0.2 and 0.1 respectively. The four surrounding sidelobes are also contoured at the 0.1 level. Due to the short integration time at each pointing, the PSF for every cubelet shows high sidelobes. After the stacking, the PSF becomes more concentrated, with lower sidelobes. \textit{Bottom:} moment 0 images before and after deconvolution in a noise-free simulation. (c): the dirty moment 0 image from stacking the convolved model images. (d): the moment 0 map of the restored image after a 100 iterations of {\sc clean} deconvolution. The two images have the same colour scale. The sidelobes are highly suppressed.}
	\label{fig:stack_4panel}
\end{figure*}

A key motivation for cubelet stacking is improvement of PSF quality. As above, upper panels in Fig~\ref{fig:stack_4panel} demonstrate how poor PSF quality for a short equatorial VLA observation can be, and how much improvement can be made after stacking. Stacking PSFs is equivalent to gaining {\it uv}-coverage in the normal manner via Earth-rotation synthesis. 
The stacking procedure concentrates the PSF, and suppresses sidelobes.

Moreover, assuming that a detection is made after the stacking procedure, the stacked image can also be deconvolved. For interferometric stacking, where the galaxies may be partially resolved, this procedure mitigates against the effects of sidelobes on the final result as demonstrated above. For the purposes of deconvolution, we use the standard CASA task {\sc deconvolve}, using a loop gain of 0.1.

\subsubsection{Noise-free cubelet stacking}

 The moment 0 images of the cubelet stacks, both before and after deconvolution, are shown in Fig~\ref{fig:stack_4panel}. Both are extracted from the central 16 channels where H\,\textsc{i} model data exist. In the right panel, 100 {\sc clean} iterations are used. The result clearly demonstrates that, despite the statistical nature of the stacks, the end result is that sidelobes are very effectively suppressed and a relatively good 2-d Gaussian image is recovered. There are pathological cases where such a technique will not work, but for a large number of objects, this simulation demonstrates that deconvolution of a stack is almost as good as for a single noise-free unstacked image.

\begin{figure}
    \includegraphics[width=\columnwidth]{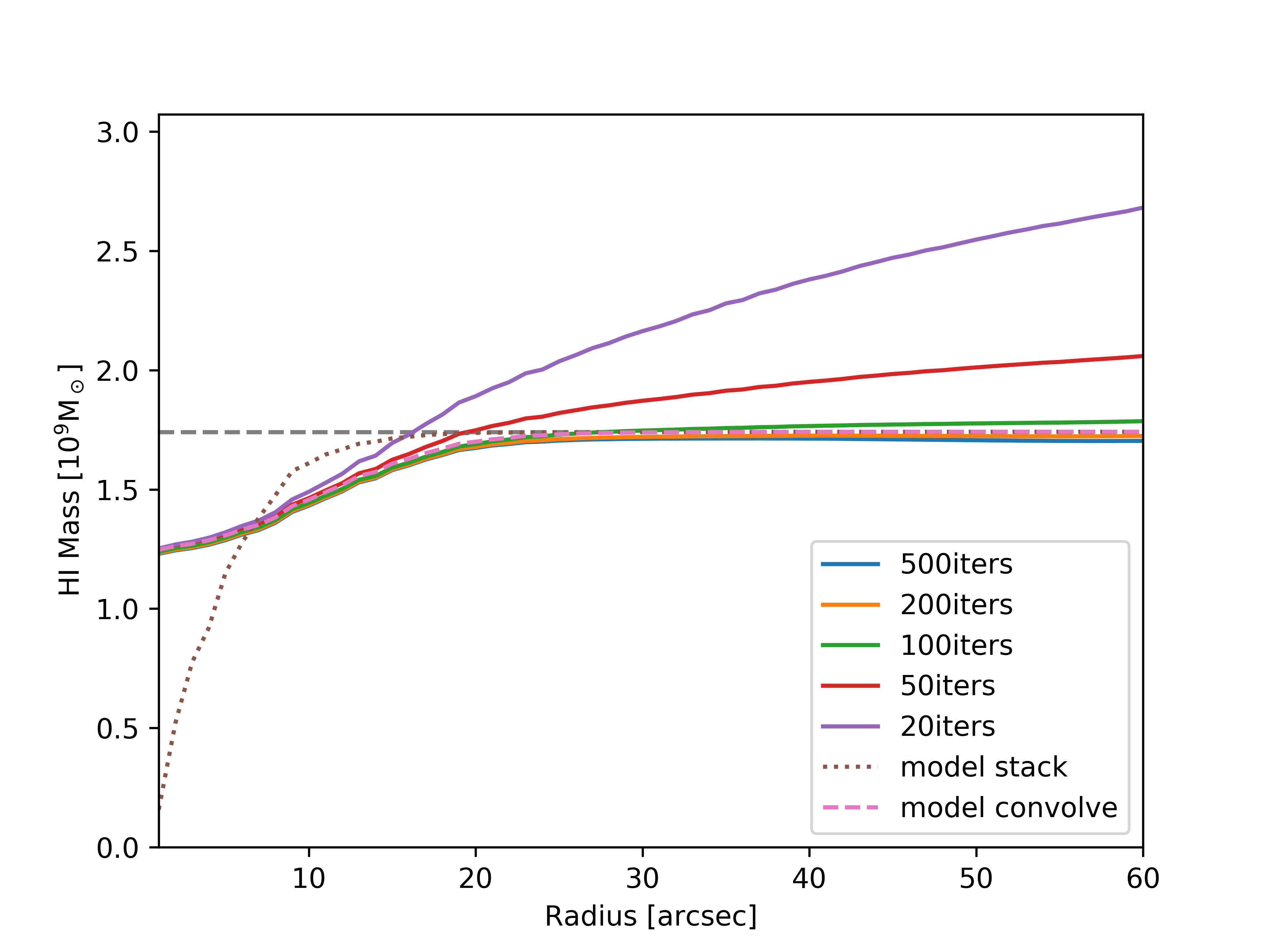}
    \caption{Integrated mass as a function of aperture size for the noise-free simulation. The {\sc clean} gain is 0.1. The five solid curves are the H\,\textsc{i} masses derived with different numbers of {\sc clean} cycles. 
    The dotted curve is the expected H\,\textsc{i} mass generated from the stacked model image, which converges to $1.74\times 10^9$ M$_\odot$ at a radius of $\sim$20$\arcsec$ (this value is also indicated by the dashed horizontal line). 
    The dashed curve is a similar result, after  convolving the stacked model directly with the Gaussian beam similar in size to the central part of the the stacked PSF. For a successful deconvolution, at least 100 iterations are needed for this noise-free simulation. 
    }
    \label{fig:curve_no-noise}
\end{figure}

The derived H\,\textsc{i} mass as a function of apertures in is shown in Fig~\ref{fig:curve_no-noise}. With a $20\arcsec$ radius circular aperture, all the flux is recovered. However, the {\sc clean} depth is important for larger apertures. For more than 100 iterations, the measured H\,\textsc{i} mass converges to the predicted value at large apertures. For less than 50 iterations, the recovered flux is $>20$\% too high at our largest measured aperture of $60\arcsec$.
The recovered H\,\textsc{i} mass for 100 iterations and an aperture of $20\arcsec$ is $\rm 1.69\times10^9M_\odot$, or about 97\% of the expected value.

\subsubsection{Cubelet stacking with noise}
\label{sec:noise}

We now make 20 realisations of noise using the method described in Section~\ref{sec:noise-cube}. As with the noise-free case, we use the standard CASA task {\sc deconvolve} with a loop gain of 0.1. We also apply our {\it a priori} expectation of the source properties via a {\sc clean} mask of $20\arcsec$ radius, and multi-scale components at 6 pixel and 12 pixel (12$\arcsec$ and $24\arcsec$, respectively).

\begin{figure}
    \includegraphics[width=\columnwidth]{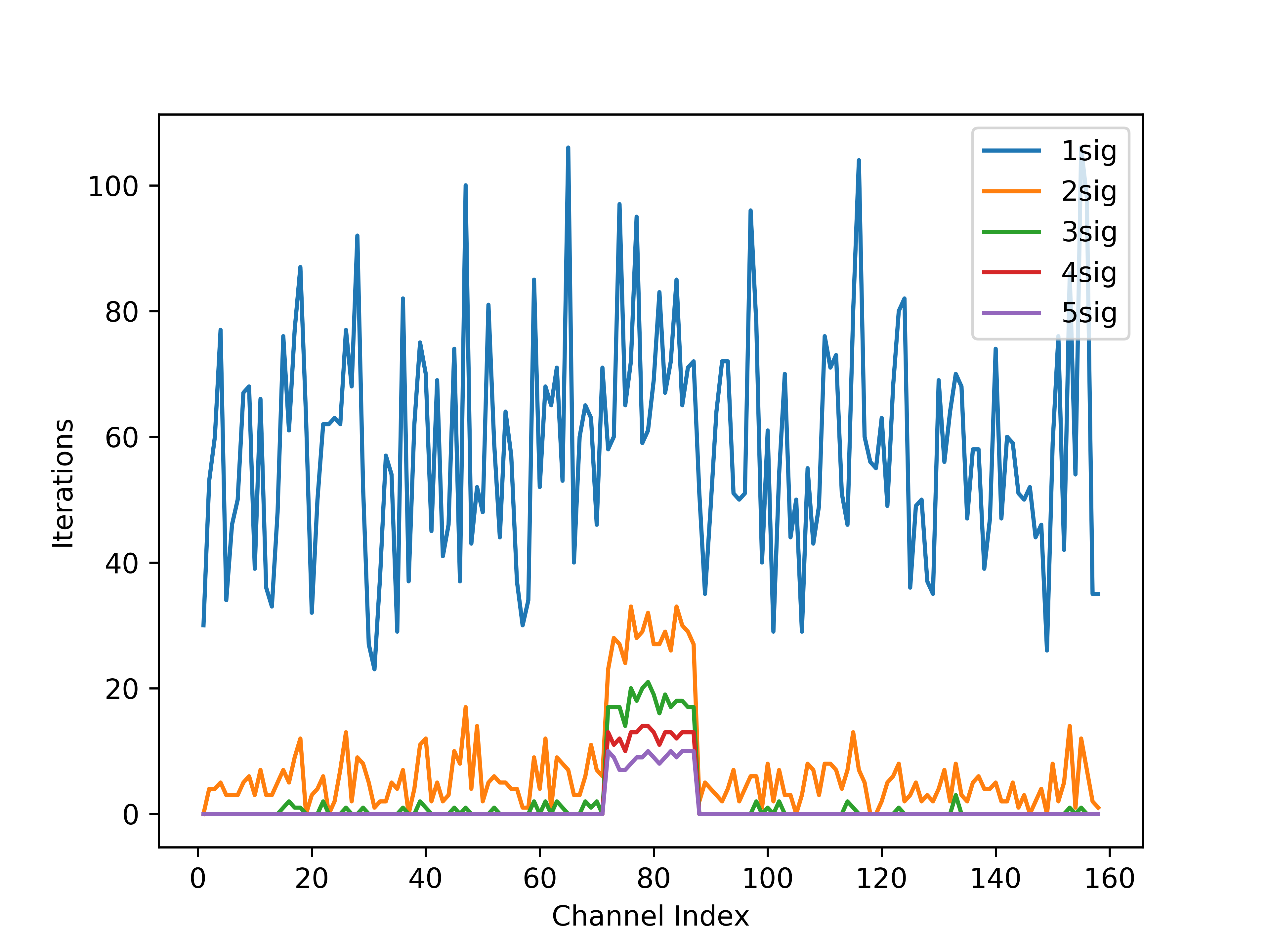}
    \caption{The number of {\sc clean} iterations as a function of channel index (i.e. frequency) for {\sc clean} depths between 1 and 5-$\sigma$. Model flux only exists in the 16 channels between 71 and 86. The {\sc clean} depth is calculated using the $rms$ calculated in the model free channels. This result is typical of the 20 noise realisations. For the 1-$\sigma$ case, channels with and without signal have similar numbers of iterations. 
    }
    \label{fig:iterations_pattern0}
\end{figure}

For a typical noise realisation with a {\sc clean} depth varying between 1 and 5-$\sigma$, the number of iterations is shown in Fig~\ref{fig:iterations_pattern0}. Even a deep {\sc clean} down to the 1-$\sigma$ level doesn't quite reach the 100 iterations suggested by the noise-free simulation. However, 
considering that we can fix the spatial integration radius at $20\arcsec$ radius, the sensitivity to the {\sc clean} threshhold should be minor (see Fig~\ref{fig:curve_no-noise}).

\begin{figure*}
    \includegraphics[width=\textwidth]{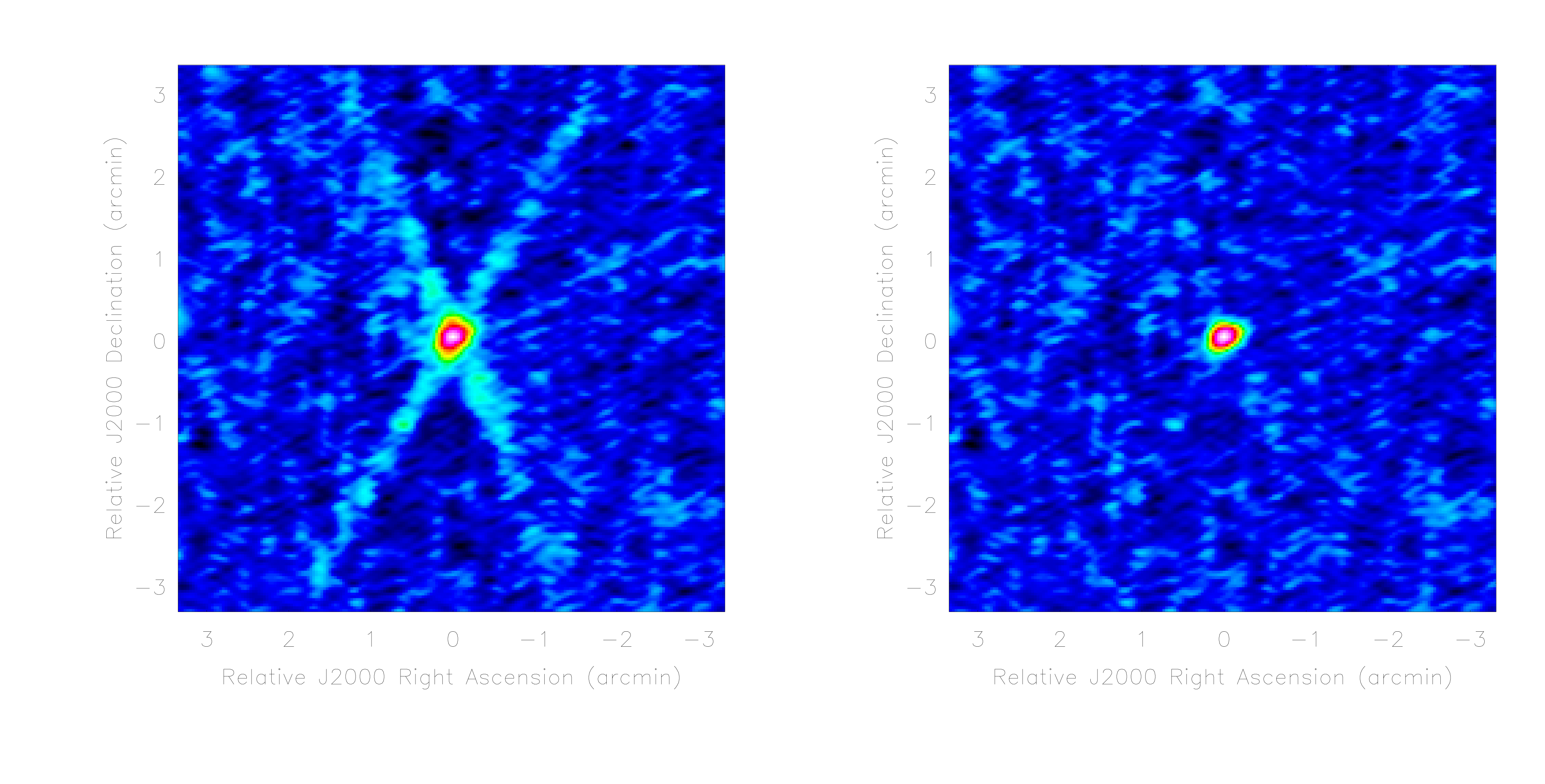}
    \caption{The moment 0 image before and after deconvolution for a typical noise realisation (\#19). Left: the dirty moment 0 image from stacking the convolved model images. Right: the restored moment 0 image after 1-$\sigma$ {\sc clean}ing. The deconvolution uses a circular cleaning mask of radius $20\arcsec$, and a multi-scale algorithm. The two plots have identical colour scales. In this simulation, the sidelobes are effectively suppressed and image quality is improved dramatically.}
    \label{fig:deconvolve-typical-noise}
\end{figure*}

\begin{figure}
    \includegraphics[width=\columnwidth]{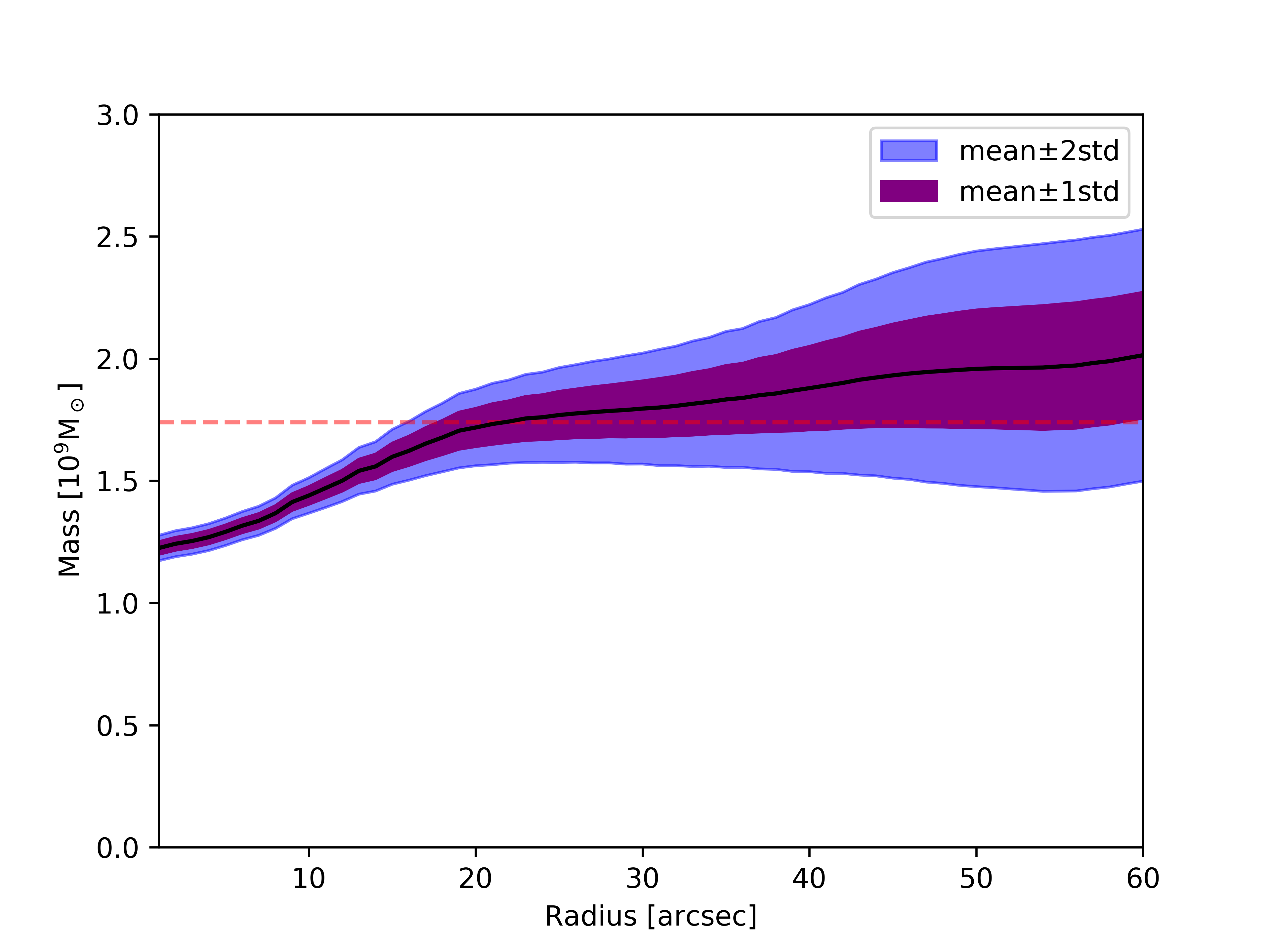}
    \caption{Integrated mass versus aperture size using a 1-$\sigma$ threshold {\sc clean}. The horizontal line is the average input mass ($1.74\times 10^9$ M$_\odot$). The solid curve show the mean value for the 20 noise realisations. The shaded regions enclose 68\% and 95\% of the realisations. At an aperture size above $\sim$20$\arcsec$ radius, the integrated masses tend to grow mildly compared with the input value, but within the errors.}
    \label{fig:shade_curve_1sigma}
\end{figure}

The moment 0 image of the central 16 channels is shown before and after 1-$\sigma$ {\sc clean}ing in Fig~\ref{fig:deconvolve-typical-noise}. The sidelobes are visibly removed and image quality is dramatically improved. 
As a function of aperture size, the average trend over the 20 realisations is shown in Fig~\ref{fig:shade_curve_1sigma} (the Appendix shows the behaviour of the individual realisations). As with the noise-free simulations, all the flux is accurately recovered at a $20\arcsec$ radius aperture. Greater than that aperture, there is gradual rise in the recovered H\,\textsc{i} mass, resulting in an overestimate of $\sim$20\% at the largest measured aperture of $60\arcsec$. This is similar to the results of the noise-free simulation when the deconvolution is imperfect ($<100$ {\sc clean} iterations).

\begin{figure}
    \includegraphics[width=\columnwidth]{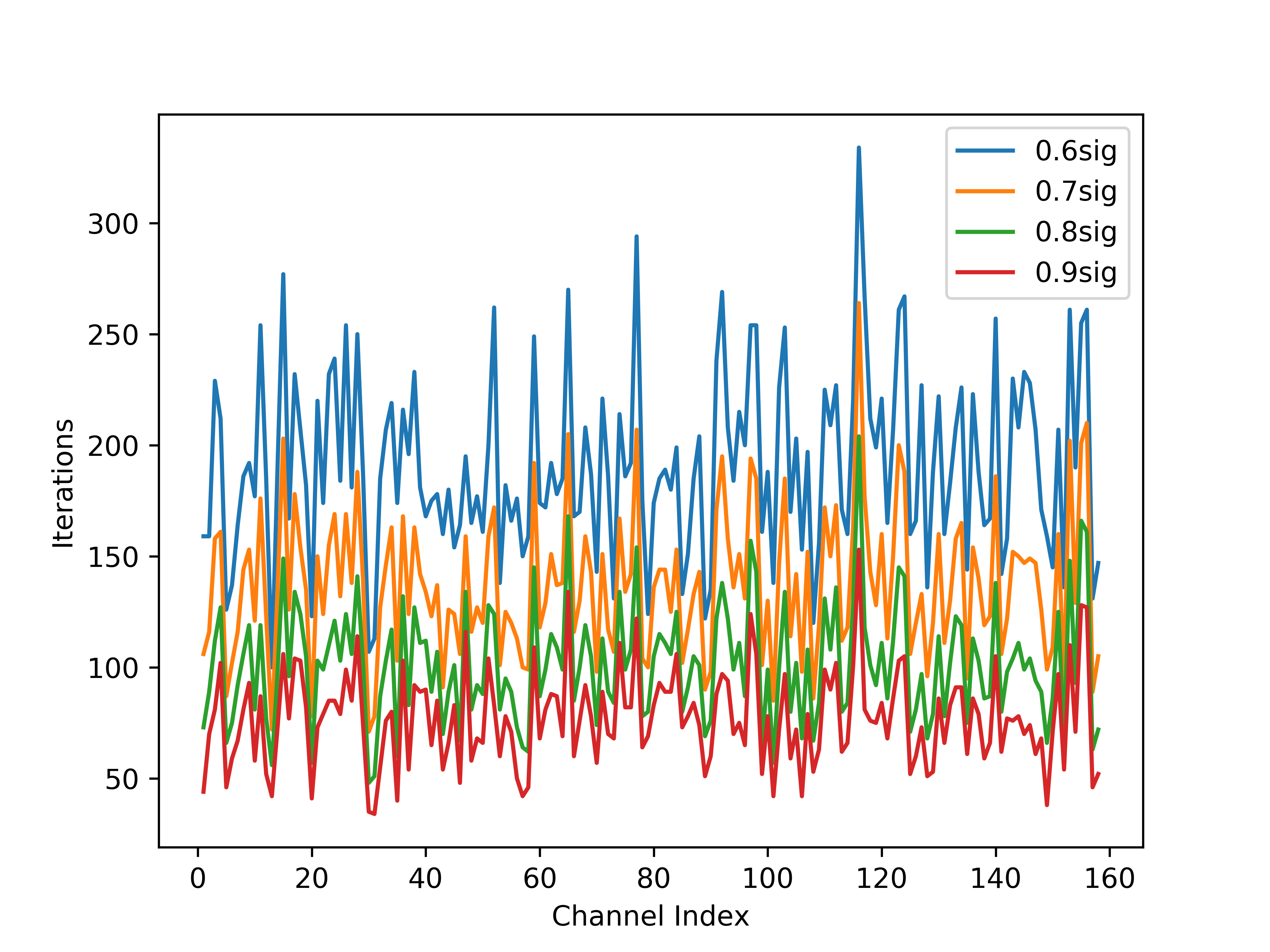}
    \caption{Number of iterations as a function of channel index (frequency) for a typical noise realisation, for {\sc clean} depths from 0.6 to 0.9-$\sigma$. In general, a depth of at least 0.7-$\sigma$ is required to reach 100 iterations.}
    \label{fig:iterations_deep}
\end{figure}

\begin{figure}
    \includegraphics[width=\columnwidth]{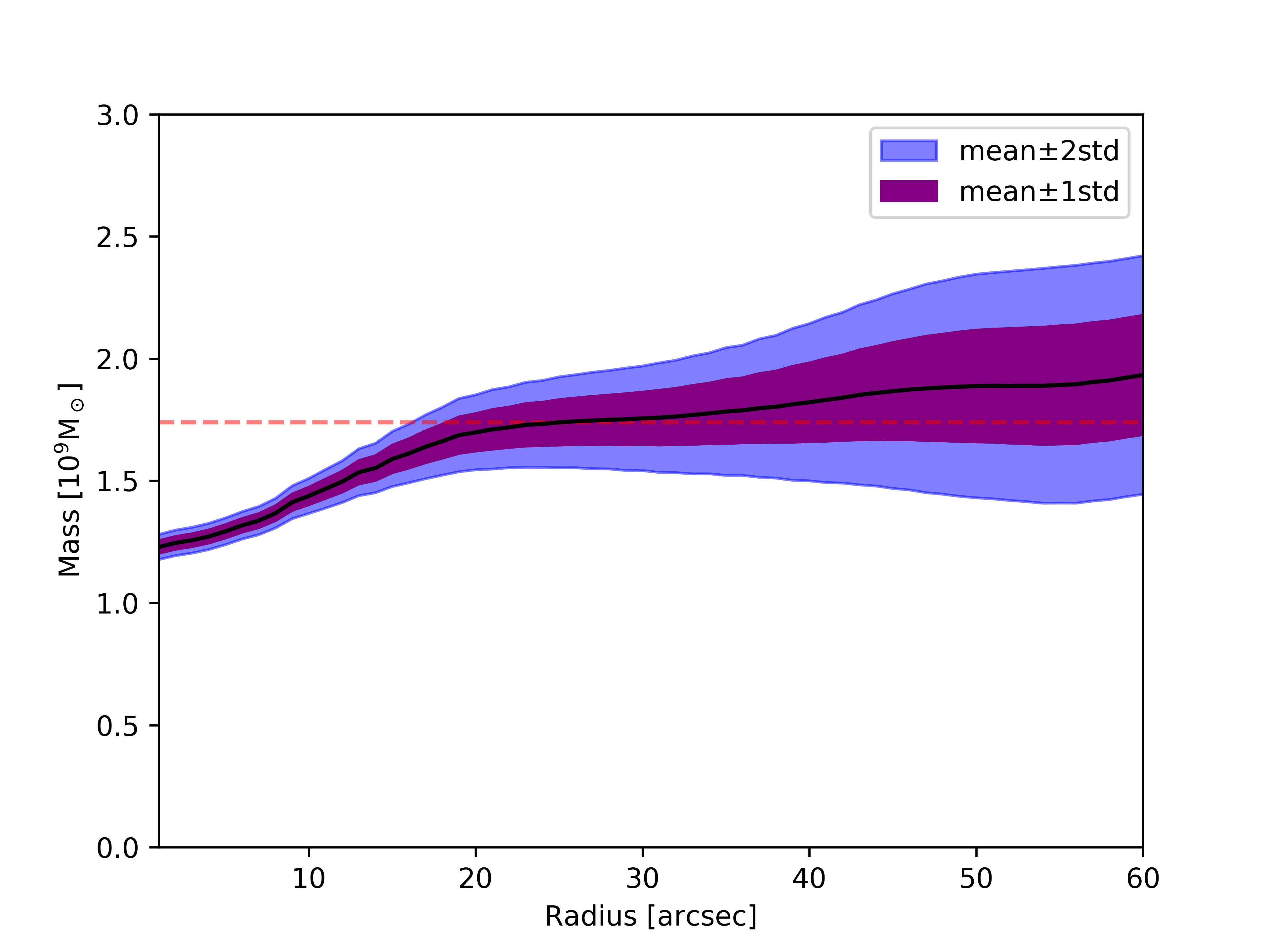}
    \caption{Integrated mass versus aperture size using a 0.7-$\sigma$ threshold {\sc clean}. The horizontal line is the average input mass ($1.74\times 10^9$ M$_\odot$). The solid curve show the mean value for the 20 noise realisations. The shaded regions enclose 68\% and 95\% of the realisations. At an aperture size above $\sim$20$\arcsec$ radius, the integrated masses tend to grow mildly compared with the input value, but within the errors.
    } 
    \label{fig:curve_0.7sigma}
\end{figure}

This demonstrates that flux can be recovered from deconvolved stacks even in the presence of noise. 
Deeper {\sc clean}ing down to 0.6 the $rms$ level was also tested. Fig~\ref{fig:iterations_deep} shows the number of {\sc clean} iterations as a function of channel index (frequency) and {\sc clean} depth. At 0.7-$\sigma$, where the average number of iterations exceeds 100, the flux--aperture relation, as with the noise-free simulations, flattens out (see Fig~\ref{fig:curve_0.7sigma}). However, there is very little difference in the result at $20\arcsec$. For the 1-$\sigma$ threshhold, the H\,\textsc{i} mass enclosed within a radius of $20\arcsec$ is $(1.72\pm 0.08) \times10^9$ M$_\odot$, about only 4\% of error around the  actual value.

\section{Discussion}

Interferometers have several advantages over single-dishes for stacking experiments. Firstly, it is often easier to reach thermal noise due to better immunity from systematic problems such as radio-frequency interference, standing waves, and gain variations. Secondly, angular resolution is usually superior, thus minimising problems that arise due to source confusion. However, the increased angular resolution often means that sources will be resolved, which in turn makes it difficult to estimate fluxes from dirty images. In fact, instruments where successful stacking experiments have been performed, such as GMRT and the VLA, have poor {\it uv}-coverage and highly non-Gaussian PSFs. These successful stacking experiments rely on lowering the resolution (by \textit{uv}-tapering) to make those extended sources unresolved.

The traditional stacking technique is to extract spectra from a galaxy's 2-d position and then stack the frequency-shifted spectra. We have introduced a new method for stacking, the so-called cubelet stacking technique. In cubelet stacking the dirty images as well as their point spread functions are stacked. The main advantages are:

1. The stacked cubelets can be deconvolved, whereas the individual cubelets cannot.

2. {\it uv}-coverage can be improved if many fields are observed over a significant length of time (as is the case in our simulation). This results in an improved PSF stack, and better deconvolution.

The method has been tested using a realistic simulation with actual noise from the DINGO-VLA survey (Chen et al. (prep)). This survey consists of 267 individual ($\sim$28~min) pointings in the equatorial GAMA 9-hr field. The combination of spatially resolved galaxies, and poor uv-coverage makes it difficult to apply the traditional spectral stacking method, as the conversion from Jy/beam to Jy is only straightforward for unresolved sources in dirty images. It is otherwise scale-dependent, in which case a Gaussian-like beam must be provided. These properties, however, make DINGO-VLA ideal for testing the viability of cubelet stacking. Success with this data set suggests that cubelet stacking will likely be much more generally applicable.

The simulation uses the same 5442 cubelet sample as defined in the DINGO-VLA survey . We firstly used scaling relations to approximate the H\,\textsc{i} sizes, shapes and masses of galaxies contained within the survey. We allocate the flux uniformly to all the pixels in 16 channels (1 MHz) of frequency. The use of top-hat distribution assumption should lead to conservative conclusions, as it populates the flux to more outskirts than reality (see discussions in Appendix~\ref{ap:exp}). We then convolve every model image with the corresponding PSF. Using this sample, we firstly apply the normal spectral stacking technique and show that the H\,\textsc{i} mass is overestimated by a factor of 2.3.

We then simulate the cubelet stacking and deconvolution in two ways.
In the noise-free simulation, we stack the convolved cubelets, then deconvolve with the stacked PSF. The sidelobes in the dirty image are highly suppressed. With a {\sc clean} that is sufficiently deep (around 100 iterations in this simulation), the actual value of H\,\textsc{i} mass is perfectly recovered, proving that in principle this method works. Cubelet stacking can reproduce the correct flux for a spatially extended stack, even where the individual PSFs are irregular and non-Gaussian.

We also conduct simulations with 20 different noise realisations, similar to the actual DINGO-VLA observational project. We randomly generate 20 sets of noise cubelets from parts of our data cubes known not to contain any emission, or catalogued galaxies, We add to the convolved cubelets, then stack. We use a multi-scale deconvolution technique, with a {\sc clean} mask over the central $20\arcsec$ radius. This allows us to apply a deep {\sc clean} where the sidelobes can be effectively suppressed. The integrated H\,\textsc{i} mass can be recovered with small uncertainty at an aperture radius of $\sim$20$\arcsec$.




Being able to deconvolve  stacks can be especially useful where stacked structure matters. For example, it should be possible to align galaxies along the optical major axis to study the stacked {\HI} diameter, and hence the relation between {\MHI} and {\DHI} as a function of redshift. With this technique the sidelobes of stacked flux in the image domain are cleaned out, enabling the detailed study of flux distributions.

For some stacking projects that observe very large sky areas to avoid cosmic variance, the limited observing time on single pointings will result in bad uv-coverage. In this case it is impossible to carry out traditional stacking analysis unless sources are unresolved. But for cubelet stacking, only the \textit{combined} uv-coverage matters. This allows observers to choose observing a large FoV while still able to apply a stacking analysis.

\section*{Acknowledgements}
\addcontentsline{toc}{section}{Acknowledgements}


Parts of this research were supported by the Australian Research Council Centre of Excellence for All Sky Astrophysics in 3 Dimensions (ASTRO 3D), through project number CE170100013, and by the Australian Research Council Centre of Excellence for All-sky Astrophysics (CAASTRO), through project number CE110001020.

\section*{Data Availability}


As described in Section 2, the simulations presented in this paper are modelled after data from the DINGO-VLA and GAMA surveys. The raw DINGO-VLA data is available from the on-line NRAO data archive\footnote{science.nrao.edu/facilities/vla/archive} under projects VLA/14B-315 and VLA/16A-341. The data for the GAMA G09 field is available from the project website\footnote{www.gama-survey.org}.


\bibliographystyle{mnras}
\bibliography{main}

\newcommand{\SortNoop}[1]{}
\begin{thebibliography}{}
\makeatletter
\relax
\def\mn@urlcharsother{\let\do\@makeother \do\$\do\&\do\#\do\^\do\_\do\%\do\~}
\def\mn@doi{\begingroup\mn@urlcharsother \@ifnextchar [ {\mn@doi@}
  {\mn@doi@[]}}
\def\mn@doi@[#1]#2{\def\@tempa{#1}\ifx\@tempa\@empty \href
  {http://dx.doi.org/#2} {doi:#2}\else \href {http://dx.doi.org/#2} {#1}\fi
  \endgroup}
\def\mn@eprint#1#2{\mn@eprint@#1:#2::\@nil}
\def\mn@eprint@arXiv#1{\href {http://arxiv.org/abs/#1} {{\tt arXiv:#1}}}
\def\mn@eprint@dblp#1{\href {http://dblp.uni-trier.de/rec/bibtex/#1.xml}
  {dblp:#1}}
\def\mn@eprint@#1:#2:#3:#4\@nil{\def\@tempa {#1}\def\@tempb {#2}\def\@tempc
  {#3}\ifx \@tempc \@empty \let \@tempc \@tempb \let \@tempb \@tempa \fi \ifx
  \@tempb \@empty \def\@tempb {arXiv}\fi \@ifundefined
  {mn@eprint@\@tempb}{\@tempb:\@tempc}{\expandafter \expandafter \csname
  mn@eprint@\@tempb\endcsname \expandafter{\@tempc}}}

\bibitem[\protect\citeauthoryear{{Baldry} et~al.,}{{Baldry}
  et~al.}{2018}]{Baldry:2018a}
{Baldry} I.~K.,  et~al., 2018, \mnras, \href
  {https://ui.adsabs.harvard.edu/abs/2018MNRAS.474.3875B} {474, 3875}

\bibitem[\protect\citeauthoryear{{Barnes} et~al.,}{{Barnes}
  et~al.}{2001}]{Barnes:2001a}
{Barnes} D.~G.,  et~al., 2001, \mnras, \href
  {https://ui.adsabs.harvard.edu/abs/2001MNRAS.322..486B} {322, 486}

\bibitem[\protect\citeauthoryear{{Bera}, {Kanekar}, {Weiner}, {Sethi}  \&
  {Dwarakanath}}{{Bera} et~al.}{2018}]{Bera:2018a}
{Bera} A.,  {Kanekar} N.,  {Weiner} B.~J.,  {Sethi} S.,   {Dwarakanath} K.~S.,
  2018, \mn@doi [\apj] {10.3847/1538-4357/aad698}, \href
  {https://ui.adsabs.harvard.edu/abs/2018ApJ...865...39B} {865, 39}

\bibitem[\protect\citeauthoryear{{Bera}, {Kanekar}, {Chengalur}  \&
  {Bagla}}{{Bera} et~al.}{2019}]{Bera:2019a}
{Bera} A.,  {Kanekar} N.,  {Chengalur} J.~N.,   {Bagla} J.~S.,  2019, \mn@doi
  [\apjl] {10.3847/2041-8213/ab3656}, \href
  {https://ui.adsabs.harvard.edu/abs/2019ApJ...882L...7B} {882, L7}

\bibitem[\protect\citeauthoryear{Bird, Garnett  \& Ho}{Bird
  et~al.}{2016}]{Bird:2016a}
Bird S.,  Garnett R.,   Ho S.,  2016, \mn@doi [Monthly Notices of the Royal
  Astronomical Society] {10.1093/mnras/stw3246}, 466, 2111

\bibitem[\protect\citeauthoryear{{Broeils} \& {Rhee}}{{Broeils} \&
  {Rhee}}{1997}]{Broeils:1997a}
{Broeils} A.~H.,  {Rhee} M.~H.,  1997, \aap, \href
  {https://ui.adsabs.harvard.edu/abs/1997A&A...324..877B} {324, 877}

\bibitem[\protect\citeauthoryear{Brown, Catinella, Cortese, Kilborn, Haynes  \&
  Giovanelli}{Brown et~al.}{2015}]{Brown:2015a}
Brown T.,  Catinella B.,  Cortese L.,  Kilborn V.,  Haynes M.~P.,   Giovanelli
  R.,  2015, \mn@doi [Monthly Notices of the Royal Astronomical Society]
  {10.1093/mnras/stv1311}, 452, 2479

\bibitem[\protect\citeauthoryear{Brown, Cortese, Catinella  \& Kilborn}{Brown
  et~al.}{2017}]{Brown:2017a}
Brown T.,  Cortese L.,  Catinella B.,   Kilborn V.,  2017, \mn@doi [Monthly
  Notices of the Royal Astronomical Society] {10.1093/mnras/stx2452}, 473, 1868

\bibitem[\protect\citeauthoryear{Catinella, Haynes, Giovanelli, Gardner  \&
  Connolly}{Catinella et~al.}{2008}]{Catinella:2008a}
Catinella B.,  Haynes M.~P.,  Giovanelli R.,  Gardner J.~P.,   Connolly A.~J.,
  2008, \mn@doi [The Astrophysical Journal] {10.1086/592328}, 685, L13

\bibitem[\protect\citeauthoryear{{Crighton} et~al.,}{{Crighton}
  et~al.}{2015}]{Crighton:2015a}
{Crighton} N. H.~M.,  et~al., 2015, \mn@doi [\mnras] {10.1093/mnras/stv1182},
  \href {https://ui.adsabs.harvard.edu/abs/2015MNRAS.452..217C} {452, 217}

\bibitem[\protect\citeauthoryear{Delhaize, Meyer, Staveley-Smith  \&
  Boyle}{Delhaize et~al.}{2013}]{Delhaize:2013a}
Delhaize J.,  Meyer M.~J.,  Staveley-Smith L.,   Boyle B.~J.,  2013, Monthly
  Notices of the Royal Astronomical Society, 433, 1398

\bibitem[\protect\citeauthoryear{{D{\'e}nes}, {Kilborn}  \&
  {Koribalski}}{{D{\'e}nes} et~al.}{2014}]{Denes:2014a}
{D{\'e}nes} H.,  {Kilborn} V.~A.,   {Koribalski} B.~S.,  2014, \mn@doi [\mnras]
  {10.1093/mnras/stu1337}, \href
  {https://ui.adsabs.harvard.edu/abs/2014MNRAS.444..667D} {444, 667}

\bibitem[\protect\citeauthoryear{{Duffy}, {Battye}, {Davies}, {Moss}  \&
  {Wilkinson}}{{Duffy} et~al.}{2008}]{Duffy:2008a}
{Duffy} A.~R.,  {Battye} R.~A.,  {Davies} R.~D.,  {Moss} A.,   {Wilkinson}
  P.~N.,  2008, \mn@doi [\mnras] {10.1111/j.1365-2966.2007.12537.x}, \href
  {https://ui.adsabs.harvard.edu/abs/2008MNRAS.383..150D} {383, 150}

\bibitem[\protect\citeauthoryear{{Duffy}, {Moss}  \& {Staveley-Smith}}{{Duffy}
  et~al.}{2012a}]{Duffy:2012b}
{Duffy} A.~R.,  {Moss} A.,   {Staveley-Smith} L.,  2012a, \mn@doi [\pasa]
  {10.1071/AS11013}, \href
  {https://ui.adsabs.harvard.edu/abs/2012PASA...29..202D} {29, 202}

\bibitem[\protect\citeauthoryear{{Duffy}, {Meyer}, {Staveley-Smith}, {Bernyk},
  {Croton}, {Koribalski}, {Gerstmann}  \& {Westerlund}}{{Duffy}
  et~al.}{2012b}]{Duffy:2012a}
{Duffy} A.~R.,  {Meyer} M.~J.,  {Staveley-Smith} L.,  {Bernyk} M.,  {Croton}
  D.~J.,  {Koribalski} B.~S.,  {Gerstmann} D.,   {Westerlund} S.,  2012b,
  \mn@doi [\mnras] {10.1111/j.1365-2966.2012.21987.x}, \href
  {https://ui.adsabs.harvard.edu/abs/2012MNRAS.426.3385D} {426, 3385}

\bibitem[\protect\citeauthoryear{{Fabello}, {Kauffmann}, {Catinella},
  {Giovanelli}, {Haynes}, {Heckman}  \& {Schiminovich}}{{Fabello}
  et~al.}{2011a}]{Fabello:2011b}
{Fabello} S.,  {Kauffmann} G.,  {Catinella} B.,  {Giovanelli} R.,  {Haynes}
  M.~P.,  {Heckman} T.~M.,   {Schiminovich} D.,  2011a, arXiv e-prints, \href
  {https://ui.adsabs.harvard.edu/abs/2011arXiv1104.0414F} {p. arXiv:1104.0414}

\bibitem[\protect\citeauthoryear{Fabello, Catinella, Giovanelli, Kauffmann,
  Haynes, Heckman  \& Schiminovich}{Fabello et~al.}{2011b}]{Fabello:2011a}
Fabello S.,  Catinella B.,  Giovanelli R.,  Kauffmann G.,  Haynes M.~P.,
  Heckman T.~M.,   Schiminovich D.,  2011b, \mn@doi [Monthly Notices of the
  Royal Astronomical Society] {10.1111/j.1365-2966.2010.17742.x}, 411, 993

\bibitem[\protect\citeauthoryear{{Fabello}, {Kauffmann}, {Catinella}, {Li},
  {Giovanelli}  \& {Haynes}}{{Fabello} et~al.}{2012}]{Fabello:2012a}
{Fabello} S.,  {Kauffmann} G.,  {Catinella} B.,  {Li} C.,  {Giovanelli} R.,
  {Haynes} M.~P.,  2012, \mn@doi [\mnras] {10.1111/j.1365-2966.2012.22088.x},
  \href {https://ui.adsabs.harvard.edu/abs/2012MNRAS.427.2841F} {427, 2841}

\bibitem[\protect\citeauthoryear{Fern{\'{a}}ndez et~al.,}{Fern{\'{a}}ndez
  et~al.}{2013}]{Fernandez:2013a}
Fern{\'{a}}ndez X.,  et~al., 2013, \mn@doi [The Astrophysical Journal]
  {10.1088/2041-8205/770/2/l29}, 770, L29

\bibitem[\protect\citeauthoryear{Fern{\'{a}}ndez et~al.,}{Fern{\'{a}}ndez
  et~al.}{2016}]{Fernandez:2016a}
Fern{\'{a}}ndez X.,  et~al., 2016, \mn@doi [The Astrophysical Journal]
  {10.3847/2041-8205/824/1/l1}, 824, L1

\bibitem[\protect\citeauthoryear{{Freudling} et~al.,}{{Freudling}
  et~al.}{2011}]{Freudling:2011a}
{Freudling} W.,  et~al., 2011, \mn@doi [\apj] {10.1088/0004-637X/727/1/40},
  \href {https://ui.adsabs.harvard.edu/abs/2011ApJ...727...40F} {727, 40}

\bibitem[\protect\citeauthoryear{{Ger{\'e}b}, {Morganti}, {Oosterloo},
  {Guglielmino}  \& {Prandoni}}{{Ger{\'e}b} et~al.}{2013}]{Gereb:2013a}
{Ger{\'e}b} K.,  {Morganti} R.,  {Oosterloo} T.~A.,  {Guglielmino} G.,
  {Prandoni} I.,  2013, \mn@doi [\aap] {10.1051/0004-6361/201322113}, \href
  {https://ui.adsabs.harvard.edu/abs/2013A&A...558A..54G} {558, A54}

\bibitem[\protect\citeauthoryear{{Giovanelli} et~al.,}{{Giovanelli}
  et~al.}{2005}]{Giovanelli:2005a}
{Giovanelli} R.,  et~al., 2005, \aj, \href
  {https://ui.adsabs.harvard.edu/abs/2005AJ....130.2598G} {130, 2598}

\bibitem[\protect\citeauthoryear{Haynes et~al.,}{Haynes
  et~al.}{2018}]{Haynes:2018a}
Haynes M.~P.,  et~al., 2018, \mn@doi [The Astrophysical Journal]
  {10.3847/1538-4357/aac956}, 861, 49

\bibitem[\protect\citeauthoryear{{Holwerda}, {Blyth}  \& {Baker}}{{Holwerda}
  et~al.}{2012}]{Holwerda:2012a}
{Holwerda} B.~W.,  {Blyth} S.~L.,   {Baker} A.~J.,  2012, in {Tuffs} R.~J.,
  {Popescu} C.~C.,  eds,  IAU Symposium Vol. 284, The Spectral Energy
  Distribution of Galaxies - SED 2011. pp 496--499 (\mn@eprint {arXiv}
  {1109.5605}), \mn@doi{10.1017/S1743921312009702}

\bibitem[\protect\citeauthoryear{{Hopkins} \& {Beacom}}{{Hopkins} \&
  {Beacom}}{2006}]{Hopkins:2006a}
{Hopkins} A.~M.,  {Beacom} J.~F.,  2006, \mn@doi [\apj] {10.1086/506610}, \href
  {https://ui.adsabs.harvard.edu/abs/2006ApJ...651..142H} {651, 142}

\bibitem[\protect\citeauthoryear{Hoppmann, Staveley-Smith, Freudling, Zwaan,
  Minchin  \& Calabretta}{Hoppmann et~al.}{2015}]{Hoppmann:2015a}
Hoppmann L.,  Staveley-Smith L.,  Freudling W.,  Zwaan M.~A.,  Minchin R.~F.,
  Calabretta M.~R.,  2015, \mn@doi [Monthly Notices of the Royal Astronomical
  Society] {10.1093/mnras/stv1084}, 452, 3726

\bibitem[\protect\citeauthoryear{Hu et~al.,}{Hu et~al.}{2019}]{Hu:2019a}
Hu W.,  et~al., 2019, \mn@doi [Monthly Notices of the Royal Astronomical
  Society] {10.1093/mnras/stz2038}, 489, 1619

\bibitem[\protect\citeauthoryear{{Johnston} et~al.,}{{Johnston}
  et~al.}{2008}]{Johnston:2008a}
{Johnston} S.,  et~al., 2008, \mn@doi [Experimental Astronomy]
  {10.1007/s10686-008-9124-7}, \href
  {https://ui.adsabs.harvard.edu/abs/2008ExA....22..151J} {22, 151}

\bibitem[\protect\citeauthoryear{{Jones}, {Papastergis}, {Haynes}  \&
  {Giovanelli}}{{Jones} et~al.}{2015}]{Jones:2015a}
{Jones} M.~G.,  {Papastergis} E.,  {Haynes} M.~P.,   {Giovanelli} R.,  2015,
  \mn@doi [\mnras] {10.1093/mnras/stv429}, \href
  {https://ui.adsabs.harvard.edu/abs/2015MNRAS.449.1856J} {449, 1856}

\bibitem[\protect\citeauthoryear{{Jones}, {Haynes}, {Giovanelli}  \&
  {Papastergis}}{{Jones} et~al.}{2016}]{Jones:2016a}
{Jones} M.~G.,  {Haynes} M.~P.,  {Giovanelli} R.,   {Papastergis} E.,  2016,
  \mn@doi [\mnras] {10.1093/mnras/stv2394}, \href
  {https://ui.adsabs.harvard.edu/abs/2016MNRAS.455.1574J} {455, 1574}

\bibitem[\protect\citeauthoryear{{Jorsater} \& {van Moorsel}}{{Jorsater} \&
  {van Moorsel}}{1995}]{Jorsater:1995a}
{Jorsater} S.,  {van Moorsel} G.~A.,  1995, \mn@doi [\aj] {10.1086/117668},
  \href {https://ui.adsabs.harvard.edu/abs/1995AJ....110.2037J} {110, 2037}

\bibitem[\protect\citeauthoryear{{Kanekar}, {Sethi}  \&
  {Dwarakanath}}{{Kanekar} et~al.}{2016}]{Kanekar:2016a}
{Kanekar} N.,  {Sethi} S.,   {Dwarakanath} K.~S.,  2016, \mn@doi [\apjl]
  {10.3847/2041-8205/818/2/L28}, \href
  {https://ui.adsabs.harvard.edu/abs/2016ApJ...818L..28K} {818, L28}

\bibitem[\protect\citeauthoryear{{Lah} et~al.,}{{Lah} et~al.}{2007}]{Lah:2007a}
{Lah} P.,  et~al., 2007, \mnras, \href
  {https://ui.adsabs.harvard.edu/abs/2007MNRAS.376.1357L} {376, 1357}

\bibitem[\protect\citeauthoryear{{Lah} et~al.,}{{Lah} et~al.}{2009}]{Lah:2009a}
{Lah} P.,  et~al., 2009, \mnras, \href
  {https://ui.adsabs.harvard.edu/abs/2009MNRAS.399.1447L} {399, 1447}

\bibitem[\protect\citeauthoryear{{Lanzetta}, {Wolfe}, {Turnshek}, {Lu},
  {McMahon}  \& {Hazard}}{{Lanzetta} et~al.}{1991}]{Lanzetta:1991a}
{Lanzetta} K.~M.,  {Wolfe} A.~M.,  {Turnshek} D.~A.,  {Lu} L.,  {McMahon}
  R.~G.,   {Hazard} C.,  1991, \mn@doi [\apjs] {10.1086/191596}, \href
  {https://ui.adsabs.harvard.edu/abs/1991ApJS...77....1L} {77, 1}

\bibitem[\protect\citeauthoryear{{Li} \& {Goldsmith}}{{Li} \&
  {Goldsmith}}{2003}]{Li:2003a}
{Li} D.,  {Goldsmith} P.~F.,  2003, \mn@doi [\apj] {10.1086/346227}, \href
  {https://ui.adsabs.harvard.edu/abs/2003ApJ...585..823L} {585, 823}

\bibitem[\protect\citeauthoryear{{Loveday} et~al.,}{{Loveday}
  et~al.}{2012}]{Loveday:2012a}
{Loveday} J.,  et~al., 2012, \mnras, \href
  {https://ui.adsabs.harvard.edu/abs/2012MNRAS.420.1239L} {420, 1239}

\bibitem[\protect\citeauthoryear{{Madau} \& {Dickinson}}{{Madau} \&
  {Dickinson}}{2014}]{Madau:2014a}
{Madau} P.,  {Dickinson} M.,  2014, \mn@doi [\araa]
  {10.1146/annurev-astro-081811-125615}, \href
  {https://ui.adsabs.harvard.edu/abs/2014ARA&A..52..415M} {52, 415}

\bibitem[\protect\citeauthoryear{{Meyer}}{{Meyer}}{2009}]{Meyer:2009a}
{Meyer} M.,  2009, in Panoramic Radio Astronomy: Wide-field 1-2 GHz Research on
  Galaxy Evolution. p.~15 (\mn@eprint {arXiv} {0912.2167})

\bibitem[\protect\citeauthoryear{{Meyer} et~al.,}{{Meyer}
  et~al.}{2004}]{Meyer:2004a}
{Meyer} M.~J.,  et~al., 2004, \mnras, \href
  {https://ui.adsabs.harvard.edu/abs/2004MNRAS.350.1195M} {350, 1195}

\bibitem[\protect\citeauthoryear{{Meyer}, {Robotham}, {Obreschkow},
  {Westmeier}, {Duffy}  \& {Staveley-Smith}}{{Meyer}
  et~al.}{2017}]{Meyer:2017a}
{Meyer} M.,  {Robotham} A.,  {Obreschkow} D.,  {Westmeier} T.,  {Duffy} A.~R.,
   {Staveley-Smith} L.,  2017, \mn@doi [\pasa] {10.1017/pasa.2017.31}, \href
  {https://ui.adsabs.harvard.edu/abs/2017PASA...34...52M} {34, 52}

\bibitem[\protect\citeauthoryear{{Nan} et~al.,}{{Nan} et~al.}{2011}]{Nan:2011a}
{Nan} R.,  et~al., 2011, \mn@doi [International Journal of Modern Physics D]
  {10.1142/S0218271811019335}, \href
  {https://ui.adsabs.harvard.edu/abs/2011IJMPD..20..989N} {20, 989}

\bibitem[\protect\citeauthoryear{{Neeleman}, {Prochaska}, {Ribaudo}, {Lehner},
  {Howk}, {Rafelski}  \& {Kanekar}}{{Neeleman} et~al.}{2016}]{Neeleman:2016a}
{Neeleman} M.,  {Prochaska} J.~X.,  {Ribaudo} J.,  {Lehner} N.,  {Howk} J.~C.,
  {Rafelski} M.,   {Kanekar} N.,  2016, \mn@doi [\apj]
  {10.3847/0004-637X/818/2/113}, \href
  {https://ui.adsabs.harvard.edu/abs/2016ApJ...818..113N} {818, 113}

\bibitem[\protect\citeauthoryear{{Noterdaeme}, {Petitjean}, {Ledoux}  \&
  {Srianand}}{{Noterdaeme} et~al.}{2009}]{Noterdaeme:2009a}
{Noterdaeme} P.,  {Petitjean} P.,  {Ledoux} C.,   {Srianand} R.,  2009, \mn@doi
  [\aap] {10.1051/0004-6361/200912768}, \href
  {https://ui.adsabs.harvard.edu/abs/2009A&A...505.1087N} {505, 1087}

\bibitem[\protect\citeauthoryear{{Noterdaeme} et~al.,}{{Noterdaeme}
  et~al.}{2012}]{Noterdaeme:2012a}
{Noterdaeme} P.,  et~al., 2012, \mn@doi [\aap] {10.1051/0004-6361/201220259},
  \href {https://ui.adsabs.harvard.edu/abs/2012A&A...547L...1N} {547, L1}

\bibitem[\protect\citeauthoryear{{Oosterloo}, {Verheijen}, {van Cappellen},
  {Bakker}, {Heald}  \& {Ivashina}}{{Oosterloo} et~al.}{2009}]{Oosterloo:2009a}
{Oosterloo} T.,  {Verheijen} M.~A.~W.,  {van Cappellen} W.,  {Bakker} L.,
  {Heald} G.,   {Ivashina} M.,  2009, in Wide Field Astronomy \&amp; Technology
  for the Square Kilometre Array. p.~70 (\mn@eprint {arXiv} {0912.0093})

\bibitem[\protect\citeauthoryear{{Prochaska}, {Herbert-Fort}  \&
  {Wolfe}}{{Prochaska} et~al.}{2005}]{Prochaska:2005a}
{Prochaska} J.~X.,  {Herbert-Fort} S.,   {Wolfe} A.~M.,  2005, \apj, \href
  {https://ui.adsabs.harvard.edu/abs/2005ApJ...635..123P} {635, 123}

\bibitem[\protect\citeauthoryear{{Rao}, {Turnshek}  \& {Nestor}}{{Rao}
  et~al.}{2006}]{Rao:2006a}
{Rao} S.~M.,  {Turnshek} D.~A.,   {Nestor} D.~B.,  2006, \mn@doi [\apj]
  {10.1086/498132}, \href
  {https://ui.adsabs.harvard.edu/abs/2006ApJ...636..610R} {636, 610}

\bibitem[\protect\citeauthoryear{{Rao}, {Turnshek}, {Sardane}  \&
  {Monier}}{{Rao} et~al.}{2017}]{Rao:2017a}
{Rao} S.~M.,  {Turnshek} D.~A.,  {Sardane} G.~M.,   {Monier} E.~M.,  2017,
  \mn@doi [\mnras] {10.1093/mnras/stx1787}, \href
  {https://ui.adsabs.harvard.edu/abs/2017MNRAS.471.3428R} {471, 3428}

\bibitem[\protect\citeauthoryear{{Rhee}, {Zwaan}, {Briggs}, {Chengalur}, {Lah},
  {Oosterloo}  \& {van der Hulst}}{{Rhee} et~al.}{2013}]{Rhee:2013a}
{Rhee} J.,  {Zwaan} M.~A.,  {Briggs} F.~H.,  {Chengalur} J.~N.,  {Lah} P.,
  {Oosterloo} T.,   {van der Hulst} T.,  2013, \mn@doi [\mnras]
  {10.1093/mnras/stt1481}, \href
  {https://ui.adsabs.harvard.edu/abs/2013MNRAS.435.2693R} {435, 2693}

\bibitem[\protect\citeauthoryear{{Rhee}, {Lah}, {Chengalur}, {Briggs}  \&
  {Colless}}{{Rhee} et~al.}{2016}]{Rhee:2016a}
{Rhee} J.,  {Lah} P.,  {Chengalur} J.~N.,  {Briggs} F.~H.,   {Colless} M.,
  2016, \mn@doi [\mnras] {10.1093/mnras/stw1097}, \href
  {https://ui.adsabs.harvard.edu/abs/2016MNRAS.460.2675R} {460, 2675}

\bibitem[\protect\citeauthoryear{{Rhee}, {Lah}, {Briggs}, {Chengalur},
  {Colless}, {Willner}, {Ashby}  \& {Le F{\`e}vre}}{{Rhee}
  et~al.}{2018}]{Rhee:2018a}
{Rhee} J.,  {Lah} P.,  {Briggs} F.~H.,  {Chengalur} J.~N.,  {Colless} M.,
  {Willner} S.~P.,  {Ashby} M. L.~N.,   {Le F{\`e}vre} O.,  2018, \mn@doi
  [\mnras] {10.1093/mnras/stx2461}, \href
  {https://ui.adsabs.harvard.edu/abs/2018MNRAS.473.1879R} {473, 1879}

\bibitem[\protect\citeauthoryear{{Songaila} \& {Cowie}}{{Songaila} \&
  {Cowie}}{2010}]{Songaila:2010a}
{Songaila} A.,  {Cowie} L.~L.,  2010, \mn@doi [\apj]
  {10.1088/0004-637X/721/2/1448}, \href
  {https://ui.adsabs.harvard.edu/abs/2010ApJ...721.1448S} {721, 1448}

\bibitem[\protect\citeauthoryear{{Verheijen}, {van Gorkom}, {Szomoru},
  {Dwarakanath}, {Poggianti}  \& {Schiminovich}}{{Verheijen}
  et~al.}{2007}]{Verheijen:2007a}
{Verheijen} M.,  {van Gorkom} J.~H.,  {Szomoru} A.,  {Dwarakanath} K.~S.,
  {Poggianti} B.~M.,   {Schiminovich} D.,  2007, \apjl, \href
  {https://ui.adsabs.harvard.edu/abs/2007ApJ...668L...9V} {668, L9}

\bibitem[\protect\citeauthoryear{{Wong} et~al.,}{{Wong}
  et~al.}{2006}]{Wong:2006a}
{Wong} O.~I.,  et~al., 2006, \mn@doi [\mnras]
  {10.1111/j.1365-2966.2006.10846.x}, \href
  {https://ui.adsabs.harvard.edu/abs/2006MNRAS.371.1855W} {371, 1855}

\bibitem[\protect\citeauthoryear{{Zafar}, {Popping}  \& {P{\'e}roux}}{{Zafar}
  et~al.}{2013}]{Zafar:2013a}
{Zafar} T.,  {Popping} A.,   {P{\'e}roux} C.,  2013, \mn@doi [\aap]
  {10.1051/0004-6361/201321153}, \href
  {https://ui.adsabs.harvard.edu/abs/2013A&A...556A.140Z} {556, A140}

\bibitem[\protect\citeauthoryear{{Zwaan}, {van Dokkum}  \& {Verheijen}}{{Zwaan}
  et~al.}{2001}]{Zwaan:2001a}
{Zwaan} M.~A.,  {van Dokkum} P.~G.,   {Verheijen} M.~A.~W.,  2001, \mn@doi
  [Science] {10.1126/science.1063034}, \href
  {https://ui.adsabs.harvard.edu/abs/2001Sci...293.1800Z} {293, 1800}

\makeatother
\end{thebibliography}


\appendix

\section{Integrated mass vs H\,\textsc{i} mass from different noise realisations}\label{app:noise}

\begin{figure*}
    \includegraphics[width=\textwidth]{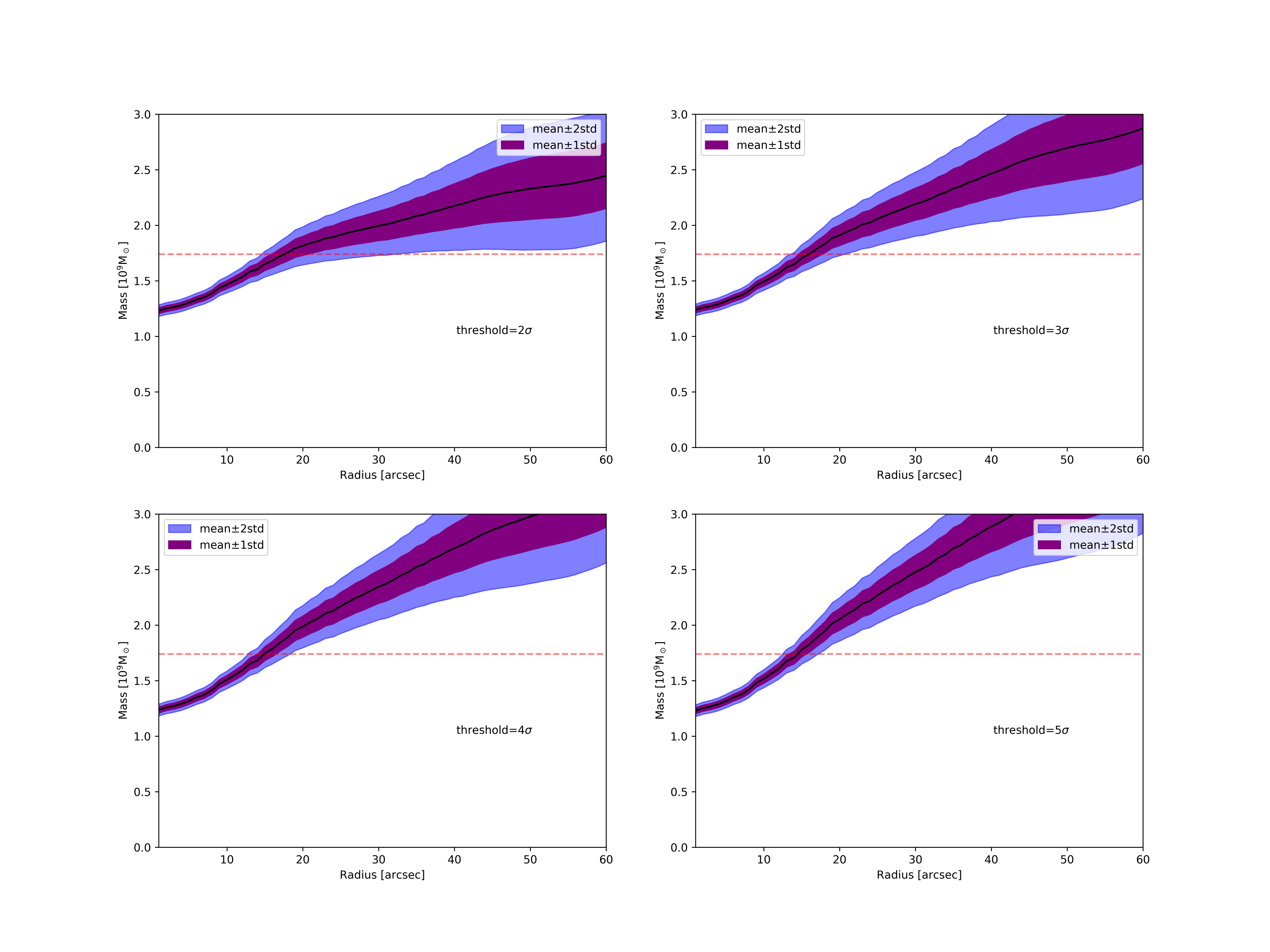}
    \caption{Similar to Fig~\ref{fig:shade_curve_1sigma}, the integrated mass versus aperture size using 2, 3, 4 and 5-$\sigma$ thresholds for {\sc clean}. The horizontal line is the average input mass ($1.74\times 10^9$ M$_\odot$). The solid curve show the mean value for the 20 noise realisations. The shaded regions enclose 68\% and 95\% of the realisations.}
    \label{fig:curv_2-5sig}
\end{figure*}

We attach the mass - aperture relation after deconvolution using a 2-5$\sigma$ threshold for all the 20 noise realisations in Fig~\ref{fig:curv_2-5sig}.

\section{Exponential and top-hat models}\label{ap:exp}

\begin{figure}
\begin{center}
    \includegraphics[width=0.95\columnwidth]{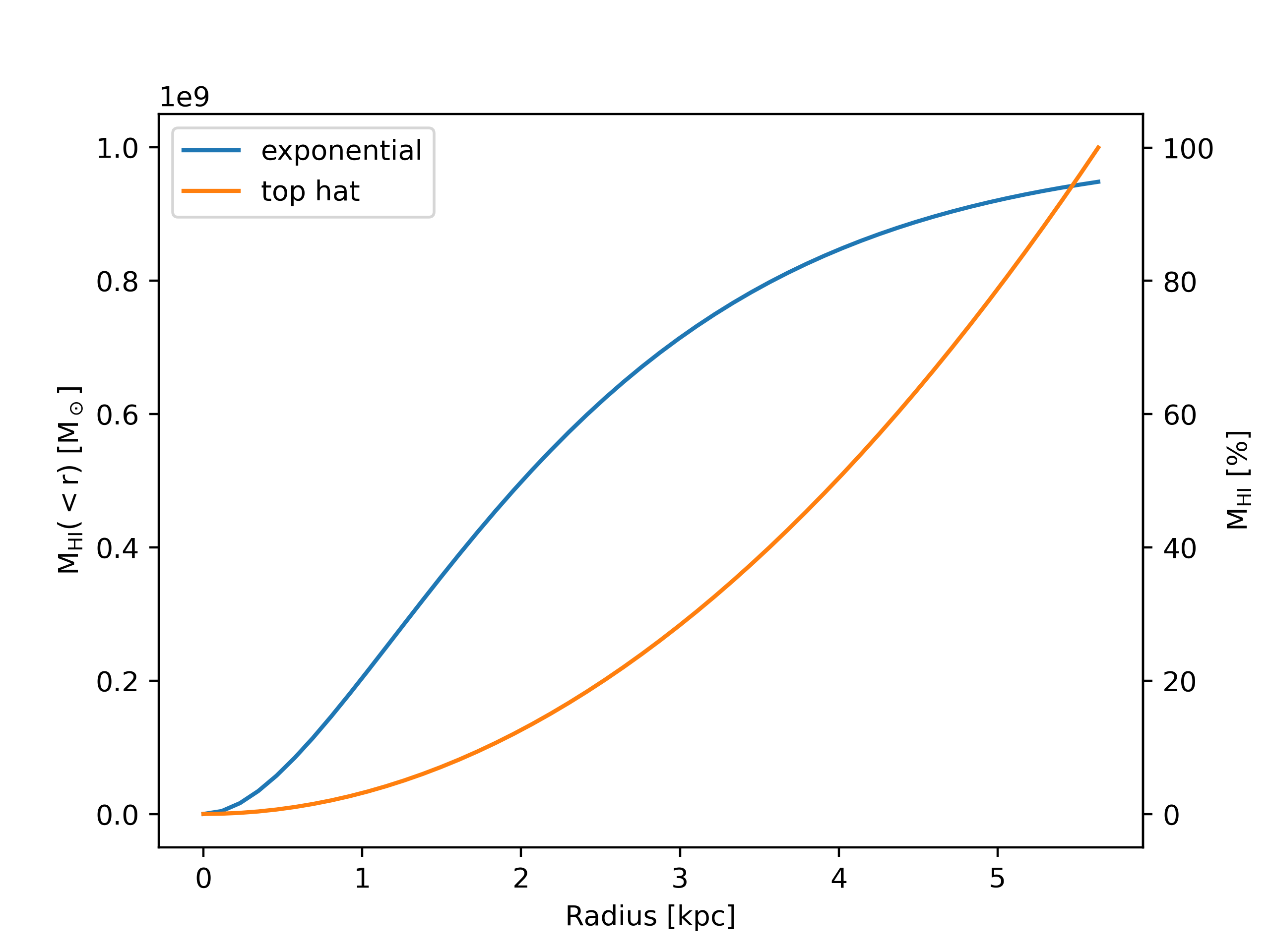}
    \caption{The integrated mass profiles from exponential and top-hat models for galaxy with ${\MHI}=10^9 {\Msun}$. The right hand $y$-axis is the percentage of the enclosed mass relative to the total {\MHI}.
    }
    \label{fig:branch-1_profiles}
\end{center}
\end{figure}

We compare top hat and exponential models for the distribution of {\HI} surface density.
The exponential model is defined by
\begin{equation} \label{eq:stellar_exp}
\Sigma(r) = \Sigma_0\exp{(-\frac{r}{r_0})} ,
\end{equation}
where the total mass is given by
\begin{equation} \label{eq:Sigma0}
\Sigma_0 = \frac{\MHI}{2\pi r_0^2}.
\end{equation}
From the previous definition of {\DHI} we solve for $r_0$ given ${\MHI}$ and $\Sigma({\RHI})=1\ {\Msun}/\rm{pc}^2$ for each of the 3622 simulated GAMA galaxies, and determine each of the surface density profiles. The integrated {\MHI} profiles for both the exponential and top-hat models for a galaxy with ${\MHI}=10^9 {\Msun}$ is shown in Fig~\ref{fig:branch-1_profiles}. The shapes of profiles change only marginally at {\MHI} values between $10^7 {\Msun}$ and $10^{10} {\Msun}$. For the exponential model, the radii that enclose 50\% and 80\% of {\MHI} are 37\% and 65\% of {\RHI}. For the top hat model, the corresponding values are 71\% and 90\% of {\RHI}. The top hat model, whilst not as realistic as the exponential model, represents a more challenging `worst-case' scenario.

\bsp	
\label{lastpage}
\end{document}